\documentclass[twocolumn]{aastex631}

\newcommand{\ar}[1]{\textcolor{red}{#1}}

\usepackage{comment}
\usepackage{float} % in preamble

\usepackage{soul}
\usepackage{gensymb}
\usepackage{amsmath}
\usepackage{{booktabs}}

\usepackage[mathlines]{lineno}
\usepackage{lipsum}  
\usepackage{orcidlink}
\usepackage{natbib}
\defcitealias{reines2020}{R20}

\graphicspath{{./}{figures/}}
\begin{document}
 
%\title{Origins of Radio Emission in Dwarf Galaxies with High-Resolution VLA Observations}
\title{Dwarf Galaxies Hosting Extreme Star-Forming Regions and (Variable) AGNs at Radio Wavelengths}
%Radio-bright Star Formation Regions and 

\author[0009-0000-4535-6340]{John-Michael Eberhard \orcidlink{0009-0000-4535-6340}}
\affiliation{eXtreme Gravity Institute, Department of Physics, Montana State University, Bozeman, MT 59717, USA}
\author[0000-0001-7158-614X]{Amy E. Reines \orcidlink{0000-0001-7158-614X}}
\affiliation{eXtreme Gravity Institute, Department of Physics, Montana State University, Bozeman, MT 59717, USA}

\begin{abstract}
{We present a detailed study of radio-detected dwarf galaxies ($M_\star \lesssim 3 \times 10^9~M_\odot$) to characterize extreme star formation and search for (variable) radio AGNs. Our sample comes from \citet{reines2020}, who used the Karl G. Jansky Very Large Array (VLA) with $\sim$0\farcs25 resolution to observe 111 dwarf galaxies with lower-resolution ($\sim$5$\arcsec$) detections in the Faint Images of the Radio Sky at Twenty Centimeters (FIRST) survey. While that work identified and focused on 13 compact radio AGN candidates in dwarf galaxies, here we focus on 16 compact radio sources consistent with star formation in dwarf galaxies. 
%In this work, we conduct a follow-up radio investigation on the remainder of the \citetalias{reines2020} sample, consisting of 1. galaxies with VLA-detected, compact radio sources that were deemed consistent with star formation in \citetalias{reines2020} and 2. galaxies lacking VLA-detected, compact radio counterparts to the FIRST detections.
We find that these objects are  %
dominated by thermal HII regions with ages $\lesssim 10$ Myr, and the most extreme sources have ionizing luminosities requiring the equivalent of $\sim$$10^{4-5}$ O-type stars. 
We also investigate the dwarf galaxies detected in FIRST but not detected in the high-resolution follow-up observations. 
%We present a follow-up investigation of dwarf galaxies %with radio detections in the Faint Images of the Radio Sky at Twenty Centimeters (FIRST) survey 
%that were observed %at 9 GHz 
%with the Karl G. Jansky Very Large Array (VLA) at $\sim$0\farcs25 resolution in \citet{reines2020}. We specifically examine the galaxies lacking compact emission consistent with active galactic nuclei (AGNs). 
%with the majority exhibiting elevated star formation surface densities that exceed the starburst intensity threshold defined by \citet{meurer1997}. % undergoing intense episodes of compact star formation. 
%Many of the host galaxies of these dense star forming regions have properties consistent with local high-redshift analogs.
%We also investigate the sources whose radio detections in the Faint Images of the Radio Sky at Twenty Centimeters (FIRST) survey  were resolved out in the high-resolution VLA data. 
Using the infrared-radio correlation parameter, we identify eight sources consistent with radio-excess AGNs. %, despite the absence of compact 9 GHz detections. %In particular 6 galaxies in particular have FIRST emission that is consistent with AGNs, using 3 different methods.
Five of these objects plus another 15 dwarf galaxies have
%We also identify 18 sources that have 
no corresponding detections in the VLA Sky Survey (VLASS) indicating variability between the FIRST and VLASS observations. The FIRST radio luminosities of these sources are significantly higher than expected for supernova-related emission, suggesting the radio variability is likely associated with AGNs. Together, these results provide new context for the presence of compact star formation and massive black holes in dwarf galaxies, and highlight the utility of radio variability and multi-resolution data for identifying the dominant power sources in low-mass systems.
%These results demonstrate that extreme star formation in dwarf galaxies can produce radio emission that mimics signatures of radio AGNs, complicating AGN identification at low galaxy masses. Moreover,
%\jme{need clearer goals for SF part -- why do I care about these objects? What do I hope to gain/ what qeustions can my research be used to answer?}
%These results indicate that dwarf galaxies are capable of hosting extraordinarily intense, localized star formation (alreday knew that), rivaling or exceeding that seen in the most luminous star-forming regions in the local universe, and offer insight into the formation of massive star clusters and the role of feedback in regulating star formation in low-mass environments (don't hae any proof of that)
}
\end{abstract}
 %{(use term ultradense hii reigon like in reines2008a? Could also look into PRSs, like in Dong+24. Thre problem is that they defined top tier PRS as those with sub-10pc scales, smoething my observations aren't high -res enough to get)} 

\section{Introduction} 
\label{sec:intro}

Radio emission provides a dust-unbiased probe of energetic processes that drive galaxy evolution, including intense star formation (SF) and active galactic nuclei (AGNs). Compact radio sources related to SF may arise from dense HII regions via thermal free--free emission, or supernova (remnants) that emit non-thermal synchrotron radiation (e.g., \citealt{condon1992}; \citealt{murphy2011}).
%Some radio emission can be indicative of intense star formation (SF); both dense HII regions (via thermal free-free emission) and supernova remnants (via non-thermal synchrotron radiation) are observed at radio frequencies \citep[e.g.,][]{condon1992, murphy2011}.
%supernovae, and supernova remnants via a combination of thermal free--free emission and non-thermal synchrotron radiation, while 
AGNs also produce radio emission, which is often dominated by synchrotron radiation related to non-thermal compact cores or collimated jets %and are often found in massive early-type galaxies 
(see \citealt{padovani2016} and \citealt{tadhunter2016}, and references therein).
%consider  Padovani 2016
Dwarf galaxies occupy a particularly informative regime for radio studies, providing insights into both extreme SF and low-mass AGNs. 

Dwarf starburst galaxies frequently host super star clusters (SSCs). These are among the densest and most massive stellar systems observed locally, with properties expected for the progenitors of globular clusters \citep[see e.g.,][]{johnson2009,
whitmore2003,oconnell2004,reines2008a, reines2008b,
weidner2010, Adamo2011}. %Many massive SSCs are found in star-forming dwarf galaxies, where SF can be particularly intense. \citep{weidner2010, Adamo2011}. %(Ho & Filippenko 1996a,b; Adamo et al. 2011).
%(vigrous SF--> SF: hunter1993, keto2005 )
%\jme{Rewords this from amy's paper?: "These extreme clusters can drastically alter the morphology of their host galaxy when the massive stars collectively explode at the end of their lives, expelling huge amounts of gas and blowing galactic-scale superbubbles (e.g., Tenorio-Tagle et al. 2007; Marlowe et al. 1995). Despite the importance of massive stars and massive clusters, their formation and earliest evolutionary stages are not well understood (Tan 2005; Zinnecker \& Yorke 2007). We do know, however, that nearly all massive stars form in clusters and that YMCs host large numbers of massive stars (Hunter 1999; Clark et al. 2005; de Wit et al. 2005). Therefore, in addition to being interesting in their own right, YMCs provide important laboratories for studying the clustered mode of massive star formation, a crucial part of understanding massive star formation in general."}
As a result, dwarf galaxies offer valuable environments for studying a clustered mode of SF that was likely prevalent in the earlier Universe and is also thought to be the dominant pathway for forming the most massive stars \citep[see][]{Zinnecker2007}.

Understanding SF in dwarf galaxies is also important for interpreting signatures that can be confused with relatively low-mass accreting massive black holes (BHs). In particular, compact HII regions and SNRs/SNe can produce radio emission with luminosities comparable to that of low-luminosity AGNs, complicating efforts to identify active massive BHs in these systems \citep[e.g.,][]{sturm2025}. A detailed characterization of SF in dwarf galaxies is therefore helpful both to study extreme SF itself and to reliably distinguish between SF- and AGN-powered radio emission in low-mass hosts.

The detailed study of massive BHs in dwarf galaxies \citep[for a review, see][]{reines22} is noteworthy because these BHs are likely to have masses between stellar mass and typical supermassive BHs ($ >10^6~M_\odot$) in massive galaxies, as seen with observed scaling relations between BH mass and host galaxy properties \citep[e.g.,][]{kor2013,mcconnell2013,reines2015,bentz2018,schutte2019}. Finding BHs in this mass regime is especially important, since they provide key constraints on BH seeding and early growth models (see \citealt{greene2020} for a review). % and references therein). 

%As a result, nearby dwarfs provide valuable laboratories for studying physical processes relevant to early galaxy evolution and feedback in low-mass systems %As such, many star-forming dwarfs serve as valuable local analogs for studying early galaxy evolution 

%Since high-redshift systems are generally faint and unresolved, examining dwarfs with intense SF to find local analogs offers a practical avenue to probe the physical processes that shaped the early universe. \ar{(More references in this paragraph.)}

Radio searches offer several advantages for identifying AGNs in dwarf galaxies. In particular, radio emission is largely unaffected by interstellar dust extinction \citep[e.g.,][]{hildebrand1983}, making it a powerful complement to optical and X-ray diagnostics, which can be weak or obscured in low-mass systems. This is especially important for detecting compact or low-luminosity AGNs that might otherwise be missed \citep[e.g.,][]{hickox2018}.
At the same time, caution is required when interpreting radio emission in dwarf galaxies, because intense SF can produce compact sources that mimic the signatures of accretion-powered activity. In addition, radio-bright AGNs are predominantly observed in massive early-type galaxies \citep[e.g.,][]{tadhunter2016}, making the detetion of of radio-bright BHs in dwarf galaxies rare.  %Such detections, however, provide a unique opportunity to probe BH accretion and feedback in the low-mass regime, to better explain why dwarfs lack radio-loud AGNs.

Radio variability provides an additional diagnostic for identifying accretion-powered activity in dwarf galaxies. AGNs are known to exhibit radio variability across a wide range of timescales due to changes in accretion rate, jet activity, or propagation effects (e.g., \citealt{nyland2020}). However, radio variability is not unique to AGNs, as transient phenomena such as supernovae (SNe), gamma-ray bursts (GRBs) and tidal disruption events (TDEs) can also produce variable or fading radio emission. As a result, variability must be interpreted in concert with luminosity and multi-epoch observations to distinguish between persistent AGN activity and transient sources.

Motivated by finding AGNs in dwarf galaxies, \citet[][\citetalias{reines2020} hereafter]{reines2020}  conducted a high-resolution ($\sim$0\farcs25) 9~GHz Very Large Array (VLA) study of 111 dwarf galaxies with radio detections in the Faint Images of the Radio Sky at Twenty Centimeters (FIRST) survey. They identified a subset of 13 compact radio sources consistent with AGN activity, which were analyzed in detail. They also identified sources with compact radio emission consistent with SF and sources that lacked compact counterparts at high angular resolution, although these were not the focus of their study. 

In this work, we present a follow-up investigation of the portion of the \citetalias{reines2020} sample that was not studied in detail in their analysis, including galaxies with SF-consistent VLA-detected compact radio sources, as well as systems lacking compact radio counterparts to the FIRST detections. We re-examine these sources using the original FIRST and high-resolution VLA observations and incorporate new data from the VLA Sky Survey (VLASS) to better characterize the nature of the radio emission. By combining spatial, luminosity, and variability information, we aim to determine the properties of the SF-consistent sources and to identify additional candidate AGNs in dwarf galaxies within sources that were not detected at high-resolution.

\section{Dwarf Galaxy Sample and Radio Data}
\label{sec:sample_dwarf_gals_radio}
Our dwarf galaxy sample is drawn from \citetalias{reines2020}, which analyzed a subset of dwarf galaxies selected from the NASA-Sloan Atlas (NSA; v0\textunderscore1\textunderscore2) {with detections in FIRST. This version of} the NSA provides redshifts, $k$-corrected stellar masses, and multi-wavelength photometry from the Sloan Digital Sky Survey (SDSS) and the Galaxy Evolution Explorer (GALEX) for galaxies at $z < 0.055$. Following \citetalias{reines2020}, galaxies with stellar masses $M_* \le 3 \times 10^9\,M_\odot$ are classified as dwarfs, and all masses and luminosities assume $h = 0.73$.

\citetalias{reines2020} obtained high-resolution 9 GHz VLA (hereafter VLA-9) observations of 111 dwarf galaxies with FIRST detections. {The detected sample of 39 galaxies} was divided into systems with secure redshifts (Sample A) and those with uncertain redshifts (Sample B). In this work, we restrict our analysis to Sample A {since these are very likely bona fide dwarf galaxies.}
{Sample A consists of 28 dwarf galaxies hosting compact VLA-9 sources.} Thirteen were classified as radio AGN candidates by \citetalias{reines2020} {and are the primary focus of that work as well as follow-up studies by \citet{sargent2022},   \citet{dong2024}, and \citet{sturm2026}. Here, we focus on the 15 dwarf galaxies exhibiting radio emission consistent with SF}, along with one borderline AGN candidate (ID~92) that was reclassified as a star-forming object based on \textit{HST} imaging that reveals a giant HII region \citep{sturm2026}. We refer to this subset as the SF sample. 

We also focus on objects that were not detected in the high-resolution follow-up observations of \citetalias{reines2020}. These non-detections may arise if the radio emission detected by FIRST is resolved out at higher angular resolution or if the source was no longer present at the time of the VLA 9~GHz observations, consistent with variable or transient radio emission. Of the 72 non-detections, we confirmed reliable redshifts (and thus stellar masses) for 59 galaxies; we discard the remaining 13 objects because their dwarf-galaxy classification cannot be robustly verified due to uncertain redshifts. We refer to this sample of 59 dwarf galaxies as the ``9~GHz non-detection'' (9GHz-ND) sample.
%discard 104 as well --> put into sample B
%Of the remaining VLA-9 targets, 59 Sample A galaxies were not detected at 9~GHz; these comprise the non-detection (ND) sample.

% -- which corresponds to a beam area $\sim400$ times smaller than FIRST—
In this work, we employ three sets of radio data: FIRST, VLA-9 and VLASS, the latter providing completely new observations for these galaxies. FIRST performed observations between 1993 and 2011 %\ar{(FIX YEAR)}
, and provides 1.4~GHz imaging with $5\arcsec$ angular resolution and a typical sensitivity of $\sim100\,\mu$Jy~beam$^{-1}$ over $\sim10{,}000$~deg$^2$ of sky \citep{white1997}. The VLA-9 observations were targeted observations conducted between February and May 2014, achieving $\sim0\farcs25$ resolution % -- which corresponds to a beam area $\sim400$ times smaller than FIRST—
and typical rms sensitivities of $\sim15\,\mu$Jy~beam$^{-1}$ \citepalias{reines2020}. VLASS observed the entire sky north of $-40^\circ$ declination over three epochs between 2017 and 2024 at 3~GHz, with an angular resolution of $2\farcs5$ and a typical rms sensitivity of $\sim120\,\mu$Jy~beam$^{-1}$ per epoch \citep{lacy2020}. We use the VLASS catalog produced by the Canadian Initiative for Radio Astronomy Data Analysis (CIRADA)\footnote{CIRADA is funded by a grant from the Canada Foundation for Innovation 2017 Innovation Fund (Project 35999), as well as by the Provinces of Ontario, British Columbia, Alberta, Manitoba and Quebec.} in collaboration with the National Radio Astronomy Observatory\footnote{The National Radio Astronomy Observatory is a facility of the National Science Foundation operated under cooperative agreement by Associated Universities, Inc.} and the Canadian Astronomy Data Centre. 

Together, FIRST and VLASS provide lower-resolution radio measurements at similar angular scales but separated by nearly two decades in time, enabling sensitivity to long-term variability at GHz frequencies. The intermediate-epoch, high-resolution VLA-9 observations probe smaller spatial scales and higher-frequency emission, allowing us to assess whether the radio emission is compact or resolved. Additionally, the inclusion of recent VLASS data provides an up-to-date view of the current radio properties of these sources and enables a multi-epoch, multi-resolution analysis that was not previously available.

%In summary, the dwarf galaxies from \citetalias{reines2020} fall into three categories: (1) systems with compact 9~GHz emission inconsistent with SF (AGN sample), (2) systems with compact 9~GHz emission consistent with thermal SF (SF sample), and (3) systems lacking compact 9~GHz detections (ND sample). In this paper, we focus on the non-AGN systems, with the SF sample presented in Section~\ref{sec:sf_sample}, the 9GHz-ND sample in Section~\ref{sec:nd}, and an investigation of radio variability using VLASS in Section~\ref{sec:variables}.

\section{VLA 9 GHz Star-Forming Sample}
\label{sec:sf_sample}
In this section, we analyze the properties of the SF sample: {the compact radio sources with VLA-9 detections in \citetalias{reines2020} and consistent with SF in their analysis} (16 dwarf galaxies including ID 92). Table \ref{tab:sf_radio} summarizes the properties of the radio sources in the SF sample, including the radio flux densities measured from the FIRST, VLASS, and VLA-9 datasets, where the FIRST and VLA-9 data are identical to those reported in \citetalias{reines2020}. 
The VLASS flux {densities} come from the Epoch 2 catalog provided by CIRADA, where available, noting that CIRADA has not made Epoch 3 publicly available at the time of publishing. For sources without values in the CIRADA catalog, we calculate their flux {densities using
CASA's \texttt{imfit} task by fitting the sources with Gaussians.} 
%in the image and integrating over the source to get the total flux \ar{density}. %{note that ID 42 has a much lower VLASS flux than FIRST flux (5x smaller)}%, although this could also be partially due to 2x better res of VLASS compared to first (Which if distrubted randomly would result in 4x less flux)}
%We note that those with CRIADA values had similar values when fit manually, so we find this method to be reliable. %at least for the SF sample
%The VLA-9 fluxes come directly from \citetalias{reines2020}, which were determined using \texttt{imfit}. 
We note that three galaxies (IDs 49, 54, and 62) exhibited multiple compact components in the VLA-9 data and these radio sources are listed separately in Table \ref{tab:sf_radio}.

%\begin{landscape}
\begin{deluxetable*}
{c|ccc|cccc|ccccc}
%\rotate
\tablecaption{SF-Consistent Radio Sources with VLA-9 Detections}
\tablewidth{0pt}
\tablehead{
\colhead{} &
\multicolumn{3}{|c|}{\shortstack{\textbf{FIRST}\\(1993--2011; 5\arcsec)}} &
\multicolumn{4}{c|}{\shortstack{\textbf{VLASS}\\(2017--2024; 2\farcs5)}} &
\multicolumn{5}{c}{\shortstack{\textbf{VLA-9}\\(2014; 0\farcs25)}} \\[6pt]
\colhead{ID} &
\colhead{$S_{1.4\ \mathrm{GHz}}$} &
\colhead{$\sigma_{1.4\ \mathrm{GHz}}$} &
\colhead{$\log(L_{1.4\ \mathrm{GHz}})$} &
\colhead{$S_{3\ \mathrm{GHz}}$} &
\colhead{$\sigma_{3\ \mathrm{GHz}}$} &
\colhead{$\log(L_{3\ \mathrm{GHz}})$} &
\colhead{$\alpha$} &
\colhead{$S_{9\ \mathrm{GHz}}$} &
\colhead{$\sigma_{9\ \mathrm{GHz}}$} &
\colhead{$\log(L_{9\ \mathrm{GHz}})$} &
\colhead{\shortstack{Major\\Axis}} &
\colhead{\shortstack{Point\\Source}}
}
\begin{comment}
\tablehead{
\colhead{ID} &
\colhead{$S_{1.4\; \mathrm{GHz}}$} &
\colhead{$\sigma_{1.4\; \mathrm{GHz}}$} &
%\colhead{Offset$_{1.4\; \mathrm{GHz}}$} &
\colhead{log($L_{1.4\; \mathrm{GHz}}$)} &
%\colhead{Offset$_{3\; \mathrm{GHz}}$} &
\colhead{$S_{3\; \mathrm{GHz}}$} &
\colhead{$\sigma_{3\; \mathrm{GHz}}$} &
\colhead{log($L_{3\; \mathrm{GHz}}$)} &
%\colhead{Offset$_{9\; \mathrm{GHz}}$} &
\colhead{$S_{9\; \mathrm{GHz}}$} &
\colhead{$\sigma_{9\; \mathrm{GHz}}$} &
\colhead{log($L_{9\; \mathrm{GHz}}$)} &
\colhead{Major Axis}&
\colhead{Pt Source}
}
\end{comment}
%see reinesA_MASTER.csv
%see vlass_cirada_astropy_casa.csv for vlass stuff
\colnumbers

\startdata
7  & 0.73 & 0.09&20.6  & $1.18^{a}$ & 0.36 & 20.8 & 0.6(0.4) & 172 & 20 & 19.9  & 158 & F \\
11  & 2.37  & 0.15&20.9  & 1.85 & 0.49 & 20.8 & -0.3(0.4) & 126 & 37 & 19.6   & 68 & T\\
30 & 2.01  & 0.23&20.9 & $3.02^{a}$ & 0.94 & 21.1 & 0.5(0.4) & 65 & 21 & 19.4    & 62 & T\\
31 & 6.78  & 0.60&21.1 & 4.41 & 0.72 & 20.8 & -0.6(0.2) & 63 & 19 & 19.0  & 28 & T\\
38 & 1.3  & 0.19&19.9  & $1.51^{a}$ & 0.51 & 19.9 & 0.2(0.5) & 259 & 10 & 19.2   & 60& F\\
42 & 3.01  & 0.33&21.3 & $0.59^{a}$ & 0.14 & 20.6  & -2.1(0.3) &66 & 23 & 19.7    & 76& T\\
43 & 2.58  & 0.17&20.8   & 3.35 & 0.5 & 20.9 &0.3(0.2) & 70 & 30 & 19.2   & 47& T\\
49a & 9.56  & 0.29&20.8  & 8.74 & 0.51 & 20.9&-0.1(0.1) & 108 & 22 & 18.9    & 28& T\\
49b &  \ldots & \ldots& \ldots & \ldots &  \ldots&\ldots& \ldots & 102 & 43 & 18.9   & 30& F\\
49c & \ldots  & \ldots& \ldots & \ldots & \ldots&\ldots& \ldots & 100 & 15 & 18.9   & 42& F\\
54a & 9.74  & 0.24& 20.5   & 5.35 & 0.35 & 20.3 &-0.8(0.1) & 458 & 64 & 19.2   &40& F\\
54b &  \ldots  & \ldots& \ldots & \ldots & \ldots&\ldots& \ldots & 340 & 30 & 19.1   & 33& F\\
54c & \ldots   & \ldots& \ldots & \ldots & \ldots&\ldots& \ldots & 103 & 30 & 18.5   &25& F\\
54d &  \ldots  & \ldots& \ldots & \ldots  & \ldots&\ldots& \ldots & 99 & 40 & 18.5    & 30& F\\
59 &  2.88  & 0.24&21.1  & $4.10^{a}$ & 1.00 & 21.2 &0.5(0.3) & 141 & 26 & 19.7   &72& F  \\
62a &  8.31  & 0.22&20.6  & 3.26 & 0.36 & 20.2 &-1.2(0.1) & 173 & 31 & 18.6  &25& F \\
62b &  \ldots & \ldots& \ldots & \ldots  & \ldots&\ldots& \ldots & 140 & 26 & 18.6   &30& F\\
62c & \ldots  & \ldots& \ldots & \ldots & \ldots&\ldots& \ldots & 75 & 28 & 18.6    & 61& T\\
72 &  1.67 &0.28& 20.2  & 2.23 & 0.33 & 20.3  &0.4(0.3) & 301 & 41 & 19.4   &38& F\\
74 & 2.75 & 0.45&19.9 & $2.52^{a}$ & 0.72 & 20.0 &-0.1(0.4) & 180 & 33 & 18.8   &25& T\\
91 &  3.69 & 0.41&19.8  & $4.86^{a}$ & 0.82 & 19.9&0.4(0.3) & 312 & 30 & 18.7   &15& T\\
92 &  4.13 & 0.21 & 21.6  & 3.19 & 0.31 & 21.5 &-0.3(0.1) & 1160 & 76 & 21.1 & 136& F\\
106 &  3.72 & 0.33&21.2  & 1.66 & 0.33 & 20.9  &-1.1(0.3) & 381 & 51 & 20.2  & 87& F    
\enddata
\tablecomments{%\ar{Indicate survey names and different resolutions and maybe years of observations in top header row.} 
Radio sources were analyzed using observations from FIRST, VLASS, and VLA-9; their years of observations and angular resolution are listed atop the table. Column 1: source identification number assigned from \citetalias{reines2020}. Column 2: radio flux density in mJy at 1.4 GHz from the FIRST catalog.  Column 3: estimated error in integrated FIRST flux  in mJy, based on local rms noise and fitted beam size of source, taken from the FIRST catalog.
%sig_1.4 = RMS*fitted_maj*fitted_min/(beam maj* beam min), with beam size assumed as 5.4'' by 5.4''
Column 4: log radio luminosity at 1.4 GHz, in W Hz$^{-1}$. Column 5: radio flux density in mJy at 3 GHz from Epoch 2 of the VLASS survey. These are taken from the CIRADA catalog, unless otherwise noted. %Those without values in the CIRADA catalog have their flux values calculated using CASA's \texttt{imfit} tool and are indicated by an asterisk. 
Column 6: error in the 3 GHz radio flux density in mJy.  Column 7: log radio luminosity at 3 GHz, in W Hz$^{-1}$. Column 8: radio spectral index, calculated using the FIRST and VLASS data. Column 9: radio flux density in $\mu$Jy at 9 GHz from the VLA-9 observations in \citetalias{reines2020}. Column 10: error in the 9 GHz radio flux density in $\mu$Jy. Column 11: log radio luminosity at 9 GHz, in W Hz$^{-1}$. Column 12: major axis of the 9 GHz source, in pc, as determined using the source fit from \texttt{imfit} and the redshift of the host galaxy (see Table \ref{tab:sf_host_gal}). Column 13: designation as a point source, as determined from \texttt{imfit}. \\
$^{a}$ Flux not provided in the CIRADA catalog. The reported value is calculated by fitting the Epoch 2 VLASS sources with Gaussians using CASA's \texttt{imfit} tool.}%Id 91 has way different flux w/ astropy (4.94) than with CASA (9.9) -- however, it's also in CIRADA w/o quality flags and between 2.5 and 5 arcsec -- this value tgives 4.46mJy, so we'll assume that one
\end{deluxetable*}
%\end{landscape}
\vspace{-\baselineskip}\vspace{-\baselineskip}
\label{tab:sf_radio}

\begin{figure*}
\includegraphics[width=\textwidth]{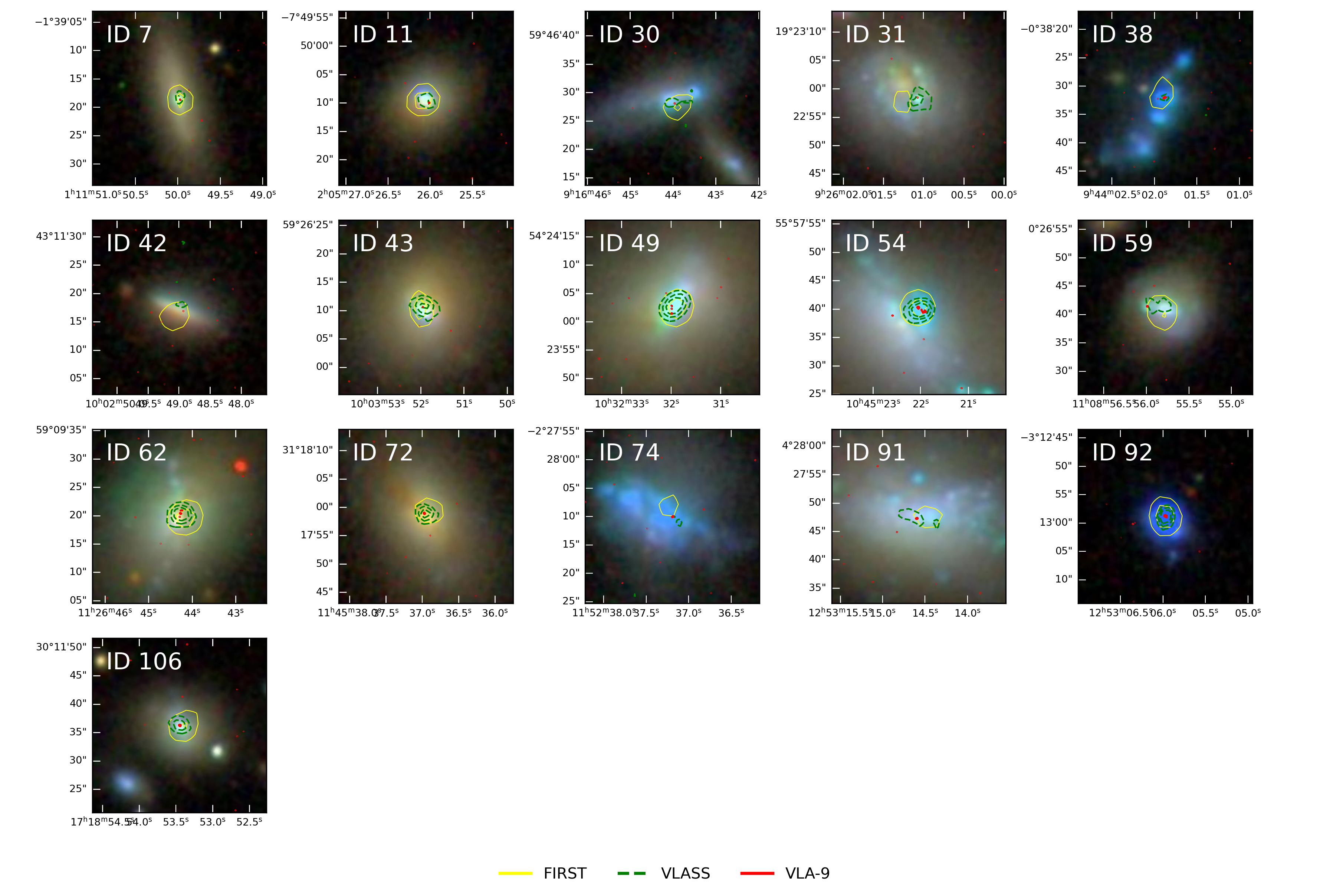}
 \caption{SDSS images of the 16 galaxies in the SF sample, with contour lines from FIRST, VLASS, and VLA-9. The contour levels for the FIRST and VLASS observations are at 3, 5, and 7$\sigma$ above the mean image flux, where $\sigma$ is the standard deviation of the background flux. The contour levels for the VLA-9 observations are for 4 and 6$\sigma$ above the mean image flux. Zoomed-in images of the VLA-9 observations are shown in Figure \ref{fig:vla9_contours}. %\ar{Can't say "etc". Be explicit and say sigma given in Table or wherever.} } %If do 3 sigma, it looks acne-esque (see id 43 in next figure) 4 sigma about the dividing line between what was classifed as source by vla9}
 }
\label{fig:optical_and_contours}
\end{figure*}

\begin{figure*}
\includegraphics[width=\textwidth]{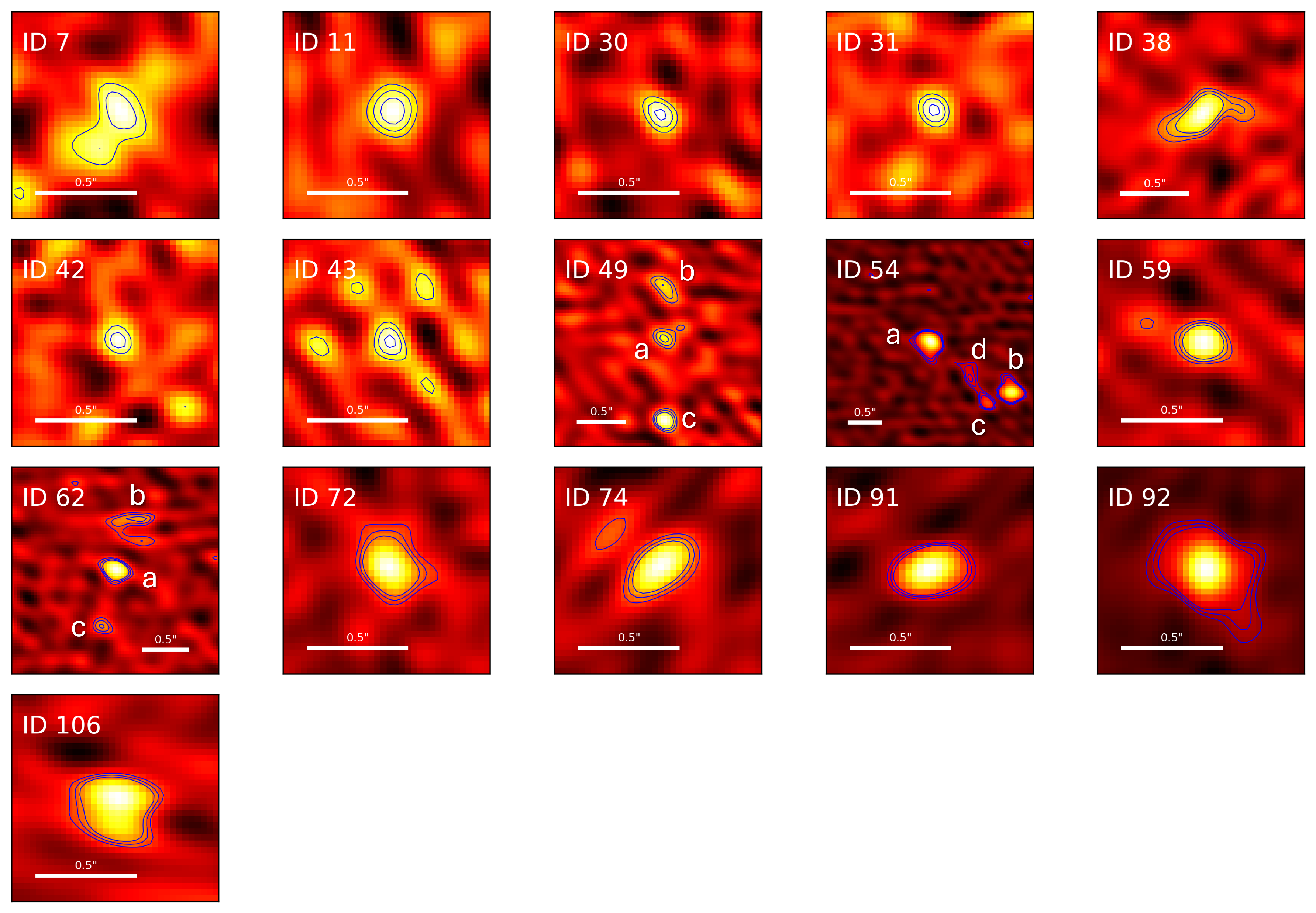}
 \caption{VLA-9 images of the 16 radio sources in the SF sample. The contour levels at 3$\sigma$, 4$\sigma$, and 5$\sigma$ above the image mean are shown, where $\sigma$ is the standard deviation of the background flux of the image. The galaxies with multiple VLA-9 sources have the separate sources labeled. %\ar{Label a, b, c etc. when multiple sources.} 
 }
\label{fig:vla9_contours}
\end{figure*}

We compare the positions of radio emission detected by FIRST, VLASS, and VLA-9 with SDSS optical images of the host galaxies in Figure~\ref{fig:optical_and_contours}.Owing to the substantially higher angular resolution of the VLA-9 data, the compact radio sources are difficult to discern in the lower-resolution FIRST and VLASS overlays. We therefore also present zoomed-in VLA-9 images highlighting the compact radio emission in Figure~\ref{fig:vla9_contours}.

%\jme{What else could I say about offset SF -- si that interesting?}
In most cases, the compact VLA-9 sources are coincident with the optical centers of their host galaxies, indicating predominantly nuclear radio emission. Two systems (IDs~31 and 59) are clear exceptions, hosting compact sources that are offset from the galaxy nucleus. In ID~31, the VLASS and VLA-9 detections are spatially coincident with each other but offset from the FIRST position; both higher-resolution detections are centered on a bright optical feature, consistent with a localized star-forming region rather than diffuse emission.

%figure from contours_ALL_92.piynb

%While these sources were still detected using imfit, since the peak VLA-9 emission is not 5$\sigma$ above the mean, it seems less likely that these are point sources (or evcen reliable?)...

%figure from vla9_contours_zoomed.ipynb

\subsection{Origin of the Compact Radio Emission}\label{sec:origin}

{In their search for radio AGNs in dwarf galaxies, \citetalias{reines2020} evaluated various possible origins for the compact radio emission, including thermal HII regions and non-thermal emission from SNRs/SNe. \citetalias{reines2020} found the radio emission to be most consistent with thermal processes from HII regions for the SF sample (e.g., see their Figure 6). However, a few sources were also consistent with luminous supernova remnants (IDs 49a/b/c, 54c/d, and 62c).} %\ar{(Are you sure these are the correct sources?)}
We further constrain the nature of the radio sources %The VLASS detections towards all 16 dwarf galaxies in our SF sample provide an additional epoch that demonstrates the persistence of the radio sources on decades-long timescales. %Here we incorporate the VLASS data to help determine if the SF is predominantly thermal or non-thermal \jme{again, r20 found they were inconsistent with non-thermal except for}.
%We use data from VLASS to substantiate the claim that the majority of the radio sources in these dwarf galaxies dominated by thermal emission. 
%, making a supernova origin unlikely. While rare radio-loud supernovae (e.g., PTF11qcj; \citealt{corsi2016}; \citealt{Palliyaguru2019}) can exhibit double-peaked radio light curves, both radio-loudness and such temporal behavior are uncommon, rendering this explanation improbable for the SF sample.
%The inclusion of VLASS data also 
by utilizing VLASS data. We estimate radio spectral indices ($\alpha$, where $S_\nu \propto \nu^\alpha$) by comparing the FIRST and VLASS flux densities (see Table \ref{tab:sf_radio}) and find that most sources exhibit spectral indices consistent with predominantly thermal emission (i.e., $\alpha \gtrsim -0.3$), while only IDs 42, 54, 62, and 106 have steep spectral indices more consistent with non-thermal emission (i.e., $\alpha \lesssim -0.8$) on arcsecond scales.  We note that at frequencies near $\sim$1~GHz, radio emission in star-forming galaxies is typica 
Publications of the Astronomical Society of Australially dominated by non-thermal processes, with average thermal fractions of order $\sim$10\% \citep{murphy2011}. The fact that most sources in our sample appear thermal even at these frequencies further supports a thermal origin. {However, we also note that the FIRST and VLASS data have different resolutions and were taken $\sim$$1-2$ decades apart so the spectral indices should be taken with caution.}

%\jme{Reorganize to not lose the fact that these are non-thermal -- megan got confused w/ other anlysis inserted in between...}
{Given our analysis and the analysis from \citetalias{reines2020}, the origin of the radio emission in the SF sample is primarily consistent with thermal bremsstrahlung (i.e., free–free emission) from ionized hydrogen in star-forming regions, while a few sources are likely dominated by synchrotron emission related to SNRs/SNe. We note that while we cannot definitively rule out radio AGNs in these galaxies, the aforementioned explanations are much more common in dwarfs.}

\begin{figure}
 \includegraphics[width=\columnwidth]{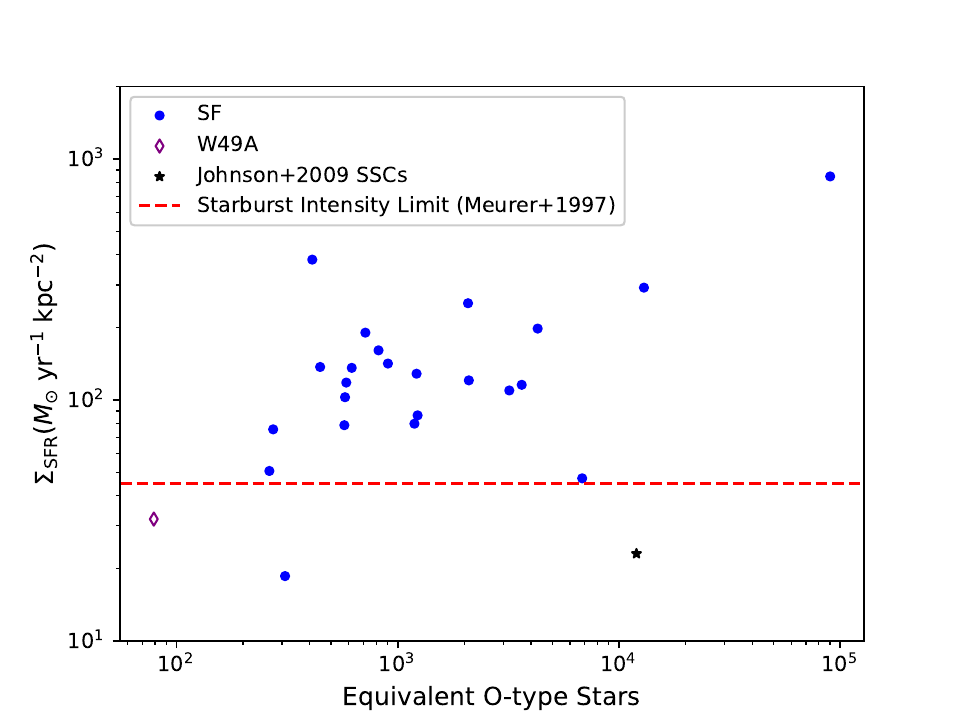}
 \caption{SFR surface density vs. expected number of typical O-type stars for the compact radio sources in the SF sample, assuming the VLA-9 radio emission is completely thermal. The Galactic region W49A and the SSCs in the BCD SBS 0335-052. \citep{johnson2009} are also shown, as well as the starburst intensity limit from \cite{meurer1997}.} %\citebf{could add reines2008a, which range from 20 to 70 o stars}}
\label{fig:sfr_vs_ostars}
\end{figure}

Under the assumption of purely thermal emission, we can characterize the radio-producing star-forming regions in the SF sample.
We estimate the Lyman-continuum production rate ($Q_{\mathrm{Lyc}}$) for the compact SF regions using the following relation from \citet{condon1992}:
\begin{equation}
\begin{aligned}
Q_{\mathrm{Lyc}} \gtrsim 6.3 \times 10^{52} 
& \left(\frac{T_e}{10^4 \; \mathrm{K}}\right)^{-0.45} 
\left(\frac{\nu}{\mathrm{GHz}}\right)^{0.1}  \\
& \times \left(\frac{L_{\nu,\mathrm{thermal}}}{10^{20} \; \mathrm{W \; Hz}^{-1}}\right)
\end{aligned}
\label{eq:qlyc}
\end{equation}
where $T_e$ is the electron temperature and $L_{\nu,\mathrm{thermal}}$ is the thermal radio luminosity. The inequality reflects the fact that some ionizing photons may be absorbed by dust. Adopting $T_e = 10^4$ K and the 9 GHz spectral luminosities from \citetalias{reines2020} (see also Table \ref{tab:sf_radio}), we derive ionizing photon production rates spanning $\log Q_{\mathrm{Lyc}} \sim 51.6$–$53.9$ s$^{-1}$. We convert to the equivalent numbers of O-type stars using the relation from \citet{vacca1996}, which provides a characteristic Lyman-continuum production rate of $Q_{\mathrm{Lyc}} = 10^{49} \,\mathrm{s^{-1}}$ for an O7.5~V star.
We find that the SF regions have a range of $\sim$400–90,000 equivalent O-type stars, {with a median of 2,000} %2087 exactly
(see Figure~\ref{fig:sfr_vs_ostars}). For comparison, W49A, one of the most luminous star-forming regions in the Milky Way, has radio emission consistent with $\sim$79 O-type stars (see \citetalias{reines2020} and references therein). Thus, even the faintest source in our sample exceeds W49A by more than a factor of five and several regions have luminosities corresponding to thousands of O-type stars, consistent with expectations for SSCs. For comparison, the blue compact dwarf (BCD) SBS 0335-052 hosts SSCs with radio emission equivalent to $\sim$12,000 O-type stars \citep{johnson2009}. We note that IDs~106 and 92 here exceed this value, with estimates of $\sim$13,000 and a whopping $\sim$90,000 O-type stars, respectively.

%fiure from histograms92.ipynb

%johnson studied low metallicty (12 + log(O/H) ≤ 7.65) galaxies --> my galaxy have metallicity estiamtes of 12 + log(O/H) between 8.1 and 8.7

%SFR SURFACE DENSITIES
We also calculate SFR surface densities ($\Sigma_{\mathrm{SFR}}$) of the radio sources using SFRs derived from the Lyman-continuum photon production rates and {areas calculated from the major and minor axes of the radio sources from VLA-9}. %\ar{(say where the areas are coming from)}. 
The SFRs are computed following the calibration of \citet{kennicutt1998}:
\begin{equation}
\label{eq:sfr_qlyc}
\mathrm{SFR} (M_\odot \; \mathrm{yr}^{-1}) = 1.08 \times 10^{-53} \; Q_{\mathrm{Lyc}}.
\end{equation}
%{compare these to known SFs}

\noindent
The resulting $\Sigma_{\mathrm{SFR}}$ values span a range of $\sim$$15- 850\,M_\odot\,\mathrm{yr^{-1}\,kpc^{-2}}$, with a median of $120\,M_\odot\,\mathrm{yr^{-1}\,kpc^{-2}}$ (see Figure~\ref{fig:sfr_vs_ostars}). Only one source (ID~62c) lies below the global starburst intensity limit of $45\,M_\odot\,\mathrm{yr^{-1}\,kpc^{-2}}$ proposed by \citet{meurer1997}. %The average $\Sigma_{\mathrm{SFR}}$ of our sample vastly exceeds that of other thermal regions reported in the literature. 
For comparison, W49A has a surface density of $\sim$32 $M_\odot\,\mathrm{yr^{-1}\,kpc^{-2}}$, and the SSCs studied by \citet{johnson2009} reach only $\sim$23 $M_\odot  \,\mathrm{yr^{-1}\,kpc^{-2}}$. %, below the \citet{meurer1997} limit. 

%\jme{So what?}
%from megan: "I really like this whole analysis! Maybe add a bit more discussion here. In my this paper (https://ui.adsabs.harvard.edu/abs/2025ApJ...979...36S/abstract) I have some citations in the discussion that might be helpful talking about highly SF dwarf galaxies. Might be cool to give this sample some contect... Sounds like htye would be really SF regions! Maybe also discuss this in the context of High z SF rates etc ? I dont know if there are any papers about that"

The combination of high inferred O-star populations and elevated $\Sigma_{\mathrm{SFR}}$ values indicates that these systems are undergoing brief, spatially concentrated bursts of SF consistent with SSCs. Therefore, they represent laboratories for investigating the physical limits of a clumpy mode of SF resembling that observed in high-z galaxies \citep[e.g.,][]{Noguchi1998,Hinojosa2016,Bournaud2016}. % \ar{(MAYBE ADD SOME REFS HERE ON CLUMPY SF AT HIGH-Z)}. 
%%(maybe {Motino-Flores2021}?). 
We note that the inferred O-star populations and $\Sigma_{\mathrm{SFR}}$ values represent upper limits under the assumption of purely thermal radio emission; any non-thermal contribution would reduce the implied ionizing photon rates and SF densities.

\subsection{Host Galaxy Properties}
\label{subsec:sf_host_gals}
In this section, we compare the host galaxy properties of dwarf galaxies containing compact, radio-bright star-forming regions (i.e., the SF sample here) to those of the broader dwarf galaxy population in the NSA. 
%This comparison is used to assess whether these systems exhibit distinct characteristics and whether their global properties are consistent with those of local analogs to high-redshift starbursts. 
A summary of the host galaxy properties for the SF sample is presented in Table~\ref{tab:sf_host_gal}.
%\subsubsection{SF Rates}
%\label{subsubsec:sf_SFRs}

\begin{figure}
\includegraphics[width=\columnwidth]{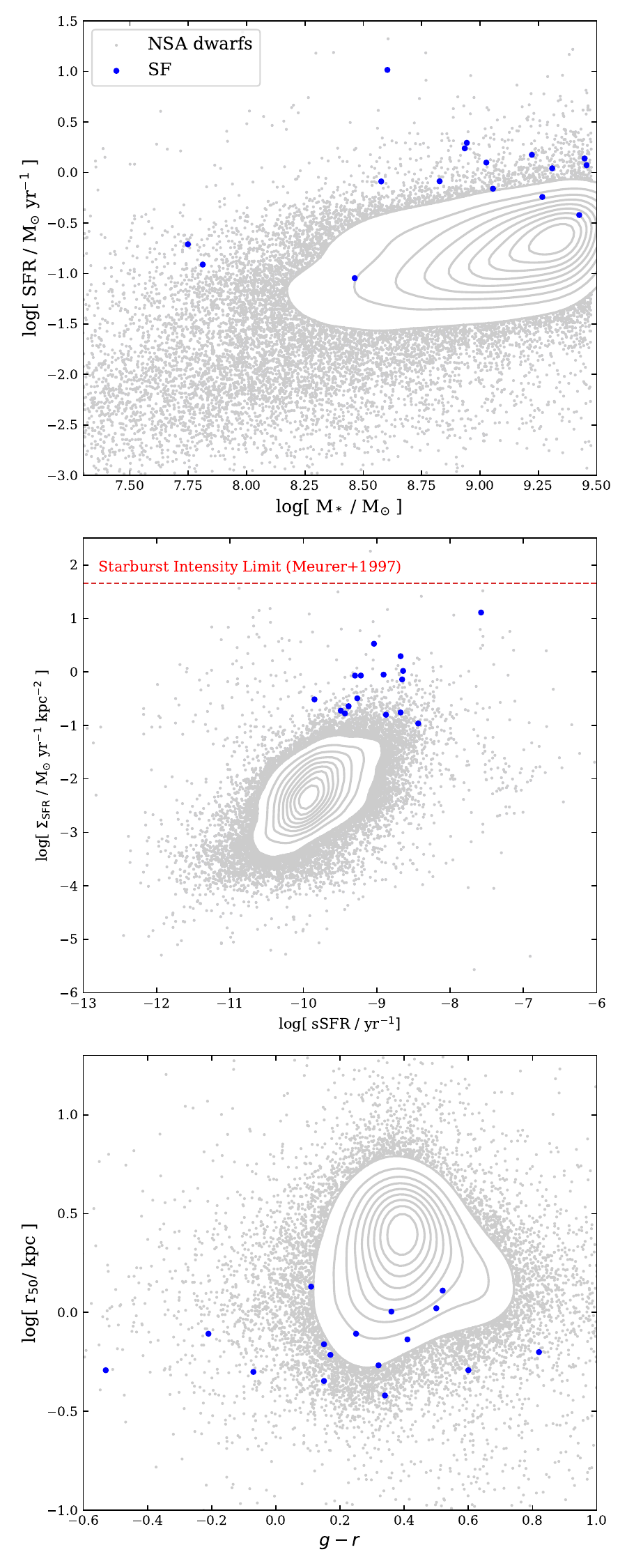}
\caption{Comparison of properties between the SF sample and the full NSA dwarf galaxy population. {Contours are plotted to enclose fixed fractions of the data, with levels corresponding to 10\%, 20\%, 30\%, etc., of the total sample. The outermost 10\% of sources, which lie farthest from the central distribution, are shown individually.} Top: SFR vs. stellar mass. Middle: SFR surface densities vs sSFRs. Bottom: half-light radius vs. $g - r$ color. }
\label{fig:dwarf_props}
\end{figure}

We first estimate galaxy-wide SFRs using the following relations from \citet{kennicutt2012} and \citet{hao2011}:
\begin{equation}
\label{eq:sfr_fuvir}
\log(\mathrm{SFR_{IR+FUV}}) = \log(L_{\mathrm{FUV, corr}}) - 43.35,
\end{equation}
where
\begin{equation}
\label{eq:fuv_correction}
L_{\mathrm{FUV, corr}} = L_{\mathrm{FUV, obs}} + 3.89 \; L_{25\mu\mathrm{m}}.
\end{equation}
Here, $L_{\mathrm{FUV, obs}}$ is the observed FUV luminosity from GALEX, and $L_{25\mu\mathrm{m}}$ is approximated from the 22 $\mu$m (W4) band luminosity measured by the Wide-field Infrared Survey Explorer (WISE; \citealt{wright2010})\footnote{This publication makes use of data products from the Wide-field Infrared Survey Explorer, a joint project of the University of California, Los Angeles, and the Jet Propulsion Laboratory/California Institute of Technology, funded by the National Aeronautics and Space Administration.}, taken from the AllWISE catalog \citep{wright2019}.  %We adopt photometry from the AllWISE catalog \citep{wright2019}, which combines data from the primary and post-cryogenic missions and provides 6–12\arcsec\ resolution imaging in four IR bands. 
Given that the ratio of the 22$\mu$m to 25$\mu$m luminosity densities is of order unity \citep{jarrett2013}, the 22$\mu$m measurements serve as reliable proxies for $L_{25\mu\mathrm{m}}$. 
All luminosities are expressed in erg s$^{-1}$, and the SFR is given in units of $M_\odot$ yr$^{-1}$. Owing to the limited spatial resolution of GALEX and WISE, these SFRs represent galaxy-wide estimates. 
When compared to the full NSA dwarf galaxy population, the SF sample is systematically offset toward higher SFRs at fixed stellar mass (top panel of Figure~\ref{fig:dwarf_props}). 
%This trend is expected given our earlier analysis, which demonstrated that the compact radio sources are powered by intense SF.

We also compare the specific SFRs (sSFR = SFR/$M_\star$) and $\Sigma_{\mathrm{SFR}}$ values of the SF sample to those of the full NSA dwarf galaxy population (middle panel of Figure~\ref{fig:dwarf_props}). The $\Sigma_{\mathrm{SFR}}$ values for the galaxies are computed using the global SFRs calculated above and half-light radii ($r_{50}$) from the NSA. The SF sample exhibits elevated sSFRs and $\Sigma_{\mathrm{SFR}}$ values relative to the broader dwarf population, although none exceed the starburst threshold defined by \citet{meurer1997}. {This suggests that while this starburst limit remains valid on galaxy-wide scales, individual compact star-forming regions can greatly exceed it (see Section \ref{sec:origin}).}

%these systems still  display elevated levels of SF, characteristic of local analogs to high-redshift galaxies.

\begin{comment}
%figure from sfrs_all_nsa92.ipynb
\begin{figure}
 \includegraphics[width=\columnwidth]{mass_sfr_dwarfs.png}
  \caption{SFR vs. stellar mass for the SF sample compared to the full NSA dwarf galaxy population. Contours denote percentile levels of the NSA distribution (10th, 20th, etc.), illustrating the relative location of the SF sample within the broader dwarf population.}

\label{fig:mass_ssfr_dwarfs}
\end{figure}

\begin{figure}
\includegraphics[width=\columnwidth]{ssfr_sigsfr_dwarfs.png}
\caption{SFR surface densities vs sSFRs for the galaxies in the SF sample compared to all dwarf galaxies in the NSA. Contours denote percentile levels of the NSA distribution (10th, 20th, etc.). }
\label{fig:ssfr_sigsfr}
\end{figure}

%from reinesA_analysis92.ipynb
\begin{figure}
\includegraphics[width=\columnwidth]{r50_gr_dwarfs.png}
\caption{Half-light radius vs. $g - r$ color for the SF sample compared to all dwarf galaxies in the NSA. Contours indicate the 10th, 20th, etc., percentiles of the NSA distribution.}
\label{fig:r50_gr}
\end{figure} 
\end{comment}

%We also analyze the color and sizes of the galaxies by comparing 
The $g - r$ colors and $r_{50}$ values of the SF sample are compared to those of all dwarf galaxies in the NSA in the bottom panel of Figure~\ref{fig:dwarf_props}. The SF sample host galaxies have $r_{50}$ values below the average radii of NSA dwarfs, indicating physically small galaxies on average. 
%perhaps a supporting a BCD-like classification. 
However, their $g-r$ colors largely overlap with the general dwarf population, indicating that intense SF may be confined to localized regions, rather than dominating the global stellar light. %Only ID 92 exhibits an especially blue color.%s, having a color consistent with the blueberry galaxy population defined by \citet{yang2017}. \ar{( say what the color cut is for the blueberries.)} %, while ID~74 may be affected by internal reddening. 

%(the most extreme star-forming galaxy in the sample) 
%92 representing the bluest case. Per the relation from yang2007, ID 92 actually has colors consistent with a blueberry galaxy.  In contrast, ID 74 exhibits a comparatively red color , which may suggest significant internal reddening or a recent decline in SF following a previous burst.
%Half-light radii in physical units (kpc) were derived from NSA half-light radii measurements in arcseconds and converted using redshift-based distances. The SF sample exhibits an even smaller average $r_{50}$ than the general dwarf population, indicating that these systems are among the more compact galaxies in the NSA catalog. This enhanced compactness supports the interpretation that these galaxies are characteristic of BCDs.

We also estimate the light-weighted ages of the stellar populations by comparing the equivalent widths (EWs) of the H$\alpha$ emission lines to \texttt{STARBURST99} population synthesis models (version 7.0.1; \citealt{Leitherer1999}) %\ar{(Need citation and model assumptions.)}. 
We adopt a metallicity of $Z = 0.004$, an instantaneous
burst of $10^4 \; M_\odot$ with a Kroupa IMF, the Geneva evolutionary
tracks with high mass loss, and the Pauldrach/Hillier atmospheres. %same parameters as reines2008a, may need to justify.  chose 0.008 for metallicity b/c 0.004 not an explicit option in models
Seven of the galaxies in the SF sample have SDSS spectra of sufficient quality that we can obtain H$\alpha$ EWs. The estimated ages for these imply very young starbursts, with inferred ages of $\sim$5.6–8.7 Myr (see Table \ref{tab:sf_host_gal}). IDs~38 and 92  have age estimates from \citet{parker2026}, with ages estimates of 13.5 and 3.8 Myr, respectively (see Table \ref{tab:sf_host_gal}). The range of ages in the SF sample are similar to those seen in other examples of extreme SF in dwarf galaxies \citep[e.g.,][]{reines2008a,reines2008b,rowland2024,sturm2025}.

Finally, we use the [NII] line strengths from available SDSS spectra to estimate the {metallicity} of the SF sample with the relation from \cite{Pettini2004}:
\begin{equation}
    12 + \mathrm{log(O/H)} = 8.90 + 0.57 \times \mathrm{log(NII/H\alpha})
\end{equation} 
While only six galaxies in the SF sample have explicit line strengths from SDSS, four additional galaxies have metallicity estimates in the literature. 
The metallicities span $12+\log(\mathrm{O/H}) \approx 7.8–8.6$, with a mean of $\sim$8.3 (see Table \ref{tab:sf_host_gal}), comparable to other BCD samples but higher than those of extremely metal-poor galaxies. %While these values are modestly elevated relative to some high-redshift analog populations, they remain subsolar and consistent with chemically evolved dwarf starbursts.

%We also estimate the metallicties of teh galaxies in our sample, using the relation from \cite{Pettini2004}
%\begin{equation}
%    12 + \mathrm{log(O/H)} = 8.90 + 0.57 \times \mathrm{log(NII/H\alpha})
%\end{equation}
%for those galaxies that have SDSS spectral lines available. We also find that 4 of the galaxies have metallicities estimated elsewhere in the literature (see Table \ref{tab:sf_host_gal}). Overall, the galaxies have metallicities between $12 + \mathrm{log(O/H)} =$ 7.83-8.62. As such, none have metallciites consistent with extremely metal-poor galaxies (XMPGs), which are defined as galaxies with metallicities 
%$12 + \mathrm{log(O/H)} \leq $ 7.65. Solar value is ($12 + \mathrm{log(O/H)} = $8.69). BCDs from Lin+2023 %Low-metallicity Galaxies from the Dark Energy Survey
%have an average oxygen abundance of $12+log(O/H)$ = 7.8
%My average is 8.3, so not especially low

At the host galaxy level, while the SF dwarfs galaxies exhibit elevated sSFRs and compact sizes relative to the broader dwarf galaxy population, they are not extreme outliers in optical color, size, or metallicity. This suggests that intense SF in dwarfs can be strongly localized, producing extreme starburst conditions without dramatically altering the global stellar light distribution.
The young inferred stellar ages further suggest that these systems trace brief, powerful episodes of SF that may play a role in shaping the evolution of low-mass galaxies. 

We note that although many of these dwarfs exhibit elevated SFRs reminiscent of high-redshift galaxies (e.g., \citealt{tolstoy2009}; \citealt{elmegreen2009})  other global properties (particularly optical color and metallicity) are generally inconsistent with typical high-redshift analogs. While a subset of the sample may share certain characteristics with early-universe systems, the absence of uniformly low metallicities and extremely blue colors makes it difficult to classify all of these galaxies as strong high-redshift analogs.

\begin{deluxetable*}
{ccccccccccccc}
%\rotate
\tablecaption{Host Galaxies of SF-Consistent 9 GHz Sources}
\tablewidth{0pt}
\tablehead{
\colhead{ID} &
\colhead{Name} &
\colhead{NSA ID} & 
\colhead{R.A.} &
\colhead{Dec.} & 
\colhead{$z$} & 
\colhead{log$(M_*)$} &
\colhead{$g-r$} &
\colhead{$r_{50}$}&
\colhead{$\mathrm{SFR_{IR+FUV}}$} &
\colhead{H${\alpha}$ Flux} &
\colhead{Age} &%FUV/IR
\colhead{12+log(O/H)}
}
%see reinesA_MASTER.csv
%see vlass_cirada_astropy_casa.csv for vlass stuff
\colnumbers
\startdata
7 & J0111-0139 & 128822 & 17.95825 & -1.65511 & 0.0158 &9.46& 0.52 & 1.29 & 2.18& --& -- & -- \\
11 &J0205-0750 &  131181 & 31.35854 & -7.83588 & 0.0126 &9.22& 0.34 & 0.38 & 2.74& -- & -- & --\\
30 &J0916+5946 &  157149 & 139.18339 & 59.77472 & 0.0143 &8.83& 0.11 & 1.35& 1.38& 12.8& 6.0 & --\\
31 &J0926+1923 &  135896 & 141.50442 & 19.38287 & 0.0091 &9.27& 0.36 & 1.01& 1.02& --& -- & --\\
38 &J0944-0038 &  7 & 146.0078 & -0.64226 & 0.0055 &7.75& -0.21 & 0.78 & 0.34& --& 13.5$^a$ & 7.83$^b$\\
42 &J1002+4311 &  157771 & 150.70413 & 43.18821 & 0.0186 &9.31& 0.5 & 1.05 & 2.03& 3.55& 7.8 & 8.49\\
43 &J1003+5926 &  27164 & 150.96616 & 59.43627 & 0.0104 &9.45& 0.41 & 0.73 & 2.47& 22.0& 7.3 & 8.62\\
49 &J1032+5424 &  34462 & 158.13323 & 54.40071 & 0.006 &8.93& 0.32 & 0.54 & 3.12& --& -- & 8.45$^c$\\
54 &J1045+5557 &  34520 & 161.34183 & 55.9612 & 0.0041 &8.58& 0.17 & 0.61& 1.47 & --&-- & --\\
59 &J1108+0026 &  906 & 167.23241 & 0.4448 & 0.0139 &9.03& 0.15 & 0.69 & 2.21& 14.6& 7.3 & 8.46\\
62 &J1126+5909 &  159841 & 171.68454 & 59.15552 & 0.0051 &9.06& 0.6 & 0.51& 1.27& 23.3&7.2 & 8.43\\
72 &J1145+3117 &  88778 & 176.40395 & 31.29955 & 0.0066 &8.43& 0.82 & 0.63& 0.70 & 6.93&8.7 & 8.57\\
74 & J1152-0228 & 3264 & 178.15497 & -2.46943 & 0.004 &7.81& -0.07 & 0.50 & 0.20& --&-- & 7.96$^c$\\
91 &J1253+0427 &  30972 & 193.31072 & 4.46323 & 0.0029 &8.46& 0.15 & 0.45& 0.14& -- &-- & --\\
92 &J1253-0312 &  3602 & 193.27489  & -3.21634 & 0.0221 &8.60& -0.53 & 0.51& 18.99&21.1$^d$ &3.8$^a$ & 8.06$^b$\\
106 &J1718+3011&  39149 & 259.72271 & 30.1934 & 0.0146 &8.94& 0.25 & 0.78& 3.61&20.8&5.6 &8.43\\    
\enddata
\tablecomments{
Column 1: identification number from \citetalias{reines2020}.
Column 2: galaxy name.
Column 3: NSA identification number.
Column 4: right ascension, in degrees.
Column 5: declination, in degrees.
Column 6: redshift.
Column 7: log galaxy stellar mass, in units of $M_\odot$.
Column 8: $g-r$ color.
Column 9: Petrosian 50\% light radius, in kpc.
Column 10: host galaxy SFR calculated using IR and FUV emission (Equation \ref{eq:sfr_fuvir}), in units of $M_\odot\,\mathrm{yr^{-1}}$.
Column 11: flux of H$\alpha$ from SDSS spectra, except where otherwise noted, in units of 10$^{-14} \mathrm{erg \;s^{-1} \; cm^{-2}}$.
Column 11: age in Myr, estimated by comparing the H$\alpha$ equivalent widths (where available in SDSS) to \texttt{STARBURST99} models, except where noted.
Column 12: metallicity based on the [N\,II] line ratio (where available), except where noted.\\
$^{a}$ Age from \citet{parker2026}\\%, who estimated stellar ages by fitting \texttt{STARBURST99} models to the stellar continuum.\\
$^{b}$ Metallicity from \citet{berg2022}\\%, derived using the [O\,II] line.\\
%$^{c}$ Metallicity from \citet{davidge1989}.\\
$^{c}$ Metallicity from \citet{Motino-Flores2021}. \\
$^{d}$ H$\alpha$ flux from \citet{sturm2026}.
%Metal Abundances UM 461 and UM 462 (Campos-Aguilar et al. 1993), Haro 02 (Davidge 1989)
%CLASSY XIII. Cutting through the Clouds – Comparing Indirect Tracers of Ionizing Photon Escape
%https://arxiv.org/html/2511.15869v1
}
\end{deluxetable*}
%\end{landscape}
\vspace{-\baselineskip}\vspace{-\baselineskip}
\label{tab:sf_host_gal}

\subsection{SF Sample in the Literature}
\label{subsec:sf_in_lit}
Several galaxies in the SF sample have been studied extensively in the literature, providing useful context for interpreting our results. IDs~38 and 92 are well-established local analogs of high-redshift star-forming galaxies, characterized by compact morphologies, low stellar masses, low metallicities, and elevated sSFRs. Both were targeted by the COS Legacy Archive Spectroscopy Survey (CLASSY), with UV spectroscopy presented by \citet{mingozzi2022} and emission-line diagnostics from \citet{mingozzi2024} confirming that their ionization is dominated by SF.

%While ID~92 hosts the highest SFR in our sample, ID~38 has one of the lowest, demonstrating that galaxies with relatively modest global SFRs can nevertheless share key properties with high-redshift analogs. This suggests that a substantial fraction of the SF sample may fall within the broader class of compact, high-redshift–like star-forming systems, even if their global SFRs vary.

IDs 49, 54, and 74 correspond to well-studied systems previously identified as blue compact dwarfs (BCDs) and high-redshift analogs.
ID 49 (Haro 2; Mrk 33; Arp 233) is a classic BCD and Wolf–Rayet galaxy that has been widely studied as a nearby analog of high-redshift starbursts (e.g., \citealt{haro1956}; \citealt{kunth1985}; \citealt{Motino-Flores2021}). %\ar{(REFS)}.
%Haro 02 also in davidge 1989, extreme blue color in thuan 1983 (this is also for Haro 3)
%Haro galaxies are a class of blue, compact galaxies, first cataloged by Mexican astronomer Guillermo Haro in 1956, known for their intense blue color, high rate of SF, and bright emission lines, often resembling "blue compact dwarfs" (BCDs) or "Green Peas"
%, including in the context of Lyman-$\alpha$ emission and ionized gas properties. 
%%It aslo was studied as Lyman-alpha emitting galaxie in Motino-Flores 2012, which has V-band imaging, UV imaging, X-ray imaging. It was also Studied as high redshift analog in Motino-Motino-Flores2021. Known BCD 
%see https://iopscience.iop.org/article/10.1086/376516/fulltext/57202.tb1.html 
ID 54 (Haro 3; Mrk 35; NGC 3353) is likewise a low-mass BCD and Wolf–Rayet system (e.g., \citealt{haro1956,steel1996}). Near-infrared (IR) and high-frequency radio imaging presented by \citet{Johnson2004} demonstrated that although localized star-forming knots are present outside the nucleus, the emission is dominated by the compact central region, which includes the components associated with our sources 54a/b/c. \citet{latimer2019} detected both radio and X-ray emission from Haro 3, with luminosities consistent with intense SF rather than AGN activity. %Johnson et al. (2004) used 8.3 and 23 GHz radio observations to study the ongoing SF in Haro 3 and detected the same radio sources presented here
ID 74 (Mrk 1307; UM 462) has similarly been classified as a BCD and examined as a local analog of high-redshift starbursts (e.g., \citealt{vanzi2003,monreal-ibero2023}).
%ID 74 is kow as Mrk 1307 in the literature. Studied as a high redshift analog in Motino-Flores2021. Known BCD %see https://iopscience.iop.org/article/10.1086/376516/fulltext/57202.tb1.html

Finally, ID~91 has been discussed in the literature as a possible AGN host based on X-ray observations. \citet{eberhard2025b} identified an X-ray source consistent with either an AGN or an ultraluminous X-ray source (ULX) based on eROSITA observations, while \citet{sacchi2024} classified the source as a high-mass X-ray binary. Importantly, the X-ray emission is significantly offset from both the galaxy nucleus and the compact radio source identified here, indicating that the X-ray and radio emission are unlikely to be physically associated.

Taken together, prior studies of galaxies in the SF sample reinforce the interpretation that these systems host extreme, compact star-forming regions rather than energetically dominant AGNs. Many of the objects have long been recognized as BCDs or local analogs of high-redshift starbursts, with multi-wavelength diagnostics consistently indicating SF as the primary source of emission. This comparison underscores that the galaxies analyzed here represent intense star-forming environments known in the nearby universe, and that some of the extreme cases have already been analyzed as galaxies that could provide a critical bridge between local starbursts and their high-redshift counterparts.

%\jme{What conclusions/discussion should I add about the SF sources?}

\section{VLA 9 GHz Non-Detection Sample}
\label{sec:nd}
{In this section, we examine the subset of dwarf galaxies from \citetalias{reines2020} with FIRST detections that lack detections in their follow-up high-resolution ($\sim$$ 0\farcs25$) VLA observations at 9 GHz. As described in Section \ref{sec:sample_dwarf_gals_radio}, we remove 13 of the 72 galaxies with unreliable redshifts and the remaining 59 dwarf galaxies are referred to as the ``9 GHz non-detection" (9GHz-ND) sample. Of these, all but 17 have subsequently been detected by VLASS (see more below). Here we aim to distinguish between radio emission consistent with SF and radio emission that requires an alternative/additional origin, such as an AGN, using the combination of FIRST and VLASS observations.%, which have similar angular resolutions of a few arcseconds.} 

%We analyze their observed properties to assess which galaxies are consistent with hosting AGNs and which exhibit radio emission attributable to SF.

%\subsection{Origin of the Radio Emission in FIRST/VLASS}

\subsection{Detections in FIRST and VLASS}
\label{subsec:nd_first_vlass}

%\ar{Notes: All have FIRST detections. Are these relative weak detections in FIRST compared to the ones detected at 9 GHz? Describe VLASS detections and measurements. How many detections. Mention 11 variable sources and refer to later sections for more details. Maybe plot of FIRST vs. VLASS flux densities for those with both, with 3 sigma sensitivities shown? \jme{avg 3 sig cutoff for VLASS was .52 mJy, avg 3 rms for FIRST was 0.44 mJy}}

Although the 9GHz–ND sources lack detections in the VLA-9 imaging, all members of this sample are detected in FIRST. There is no clear distinction in the FIRST flux density distributions between the 9GHz-ND and SF populations (see Figure~\ref{fig:nd_first_vlass}), indicating that a lower FIRST flux density alone does not explain the non-detections with the higher resolution 9 GHz observations.

Among the 42 sources detected in both FIRST and VLASS, we compute radio spectral indices using the integrated flux densities from the two surveys. {Eight sources exhibit steep spectra ($\alpha \lesssim -0.8$), while 24 display relatively flat spectra ($\alpha \gtrsim -0.3$). The remaining 10 sources fall between these regimes (see Figure~\ref{fig:nd_first_vlass}). If the emission is associated with SF, steeper indices are generally consistent with non-thermal synchrotron radiation, whereas flatter indices may indicate a larger thermal contribution from HII regions. However, AGNs can produce either flat or steep spectra.} %and therefore cannot be classified as thermal or non-thermal on the basis of spectral index alone. 
We also caution that these spectral index estimates are subject to systematic uncertainties, as FIRST and VLASS differ in angular resolution and observing epoch.

%We note that three of the sources with nominally thermal indices show unusually high VLASS flux densities ($>10$ mJy) based on \texttt{imfit} measurements and are treated as outliers \ar{(probably flat spectrum AGN)}. 

% from FIRST_fluxes.ipynb
\begin{figure}
 \includegraphics[width=\columnwidth]{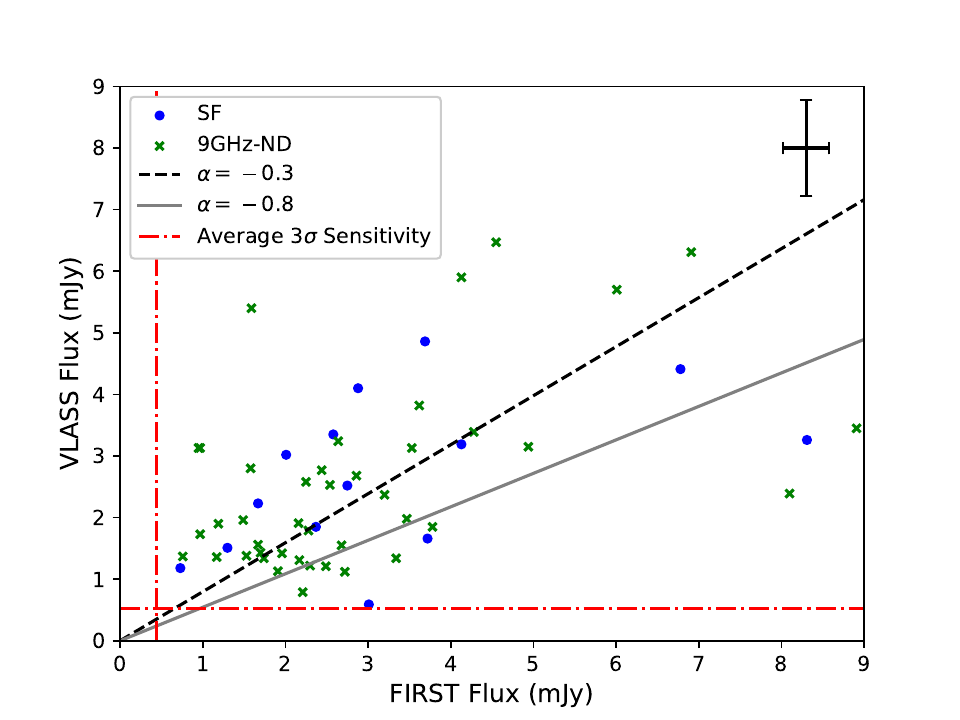}
  \caption{VLASS vs. FIRST flux densities for sources in the SF and 9GHz–ND samples with VLASS detections. Black and gray lines indicate spectral indices of $\alpha = -0.3$ and $\alpha = -0.8$, respectively. {The distribution indicates a mixture of flat- and steep-spectrum sources.} Red lines denote the average $3\sigma$ sensitivity limits for FIRST and VLASS. The mean error of the fluxes are shown in the upper right corner.}
\label{fig:nd_first_vlass}
\end{figure}

\subsection{Variable Sources}
\label{subsec:variables}

Seventeen sources in the 9GHz–ND sample (IDs 1, 12, 13, 15, 22, 29, 36, 37, 44, 45, 63, 71, 75, 78, 81, 94, and 101) are robustly detected in FIRST but lack detectable counterparts in VLASS, based on two-dimensional Gaussian fitting with CASA’s \texttt{imfit} routine and a $3\sigma$ detection threshold in VLASS, where $\sigma$ denotes the background noise of the image. Because FIRST and VLASS have comparable angular resolutions but were obtained at different epochs, resolution effects alone cannot account for the presence of a clear FIRST detection and a non-detection in VLASS. Instead, this implies a change in the source flux density between the two surveys, motivating an interpretation of intrinsic radio variability.  %We further analyze these sources in Section \ref{sec:variables}.
%We search for radio variability by comparing flux densities measured in FIRST and VLASS across the SF, 9GHz-ND, and AGN samples \ar{(what AGN samples?)}. In contrast to the higher-resolution VLA-9 data, FIRST and VLASS have comparable angular resolution and similar observing frequencies, enabling a more direct and uniform assessment of long-term radio flux evolution on decade timescales.
%Evidence for variability is found for 17 sources in the 9GHz-ND sample: IDs 1, 12, 13, 15, 22, 29, 36, 37, 44, 45, 63, 71, 75, 78, 81, 94, and 101. %All are robustly detected in FIRST but lack corresponding detections in VLASS (see Section \ref{subsec:nd_first_vlass}). \ar{(Are there any in the other samples? If not, say clearly that all variable sources are in ND sample. )} 
Representative FIRST and VLASS images for four of the 17 variable sources are shown in Figure~\ref{fig:vari_first_vlass}. 
%We investigate these objects as candidate variable sources in order to constrain the physical mechanisms responsible for their radio emission. 

Given the temporal baseline between FIRST and VLASS, several astrophysical processes could plausibly drive the observed variability, including SNe/SNRs, GRB afterglows, TDEs, or AGNs. Among these possibilities, AGNs represent the most common source of non–SF–related radio emission in galaxies, constituting a substantial fraction of the radio sky \citep[e.g.,][]{condon1998, Padovani2015, smolcic2017, Hardcastle2019}. 
%As a persistent and energetically powerful population, 
AGNs therefore provide a natural explanation for  radio variability, although alternative transient scenarios remain viable and are explored below.

\begin{comment}
\begin{figure*}
\includegraphics[width=\textwidth]{variables.png}
 \caption{FIRST and VLASS images of the eleven 9GHz-ND that had detections in FIRST but not in VLASS. The FIRST images are on the left and the VLASS images are on the right. The ellipse fit to the source in FIRST is overlaid on both the FIRST and VLASS images.} %If do 3 sigma, it looks acne-esque (see id 43 in next figure) 4 sigma about the dividing line between what was classifed as source by vla9}
\label{fig:vari_first_vlass}
\end{figure*}
\end{comment}

\begin{figure*}
\includegraphics[width=\textwidth]{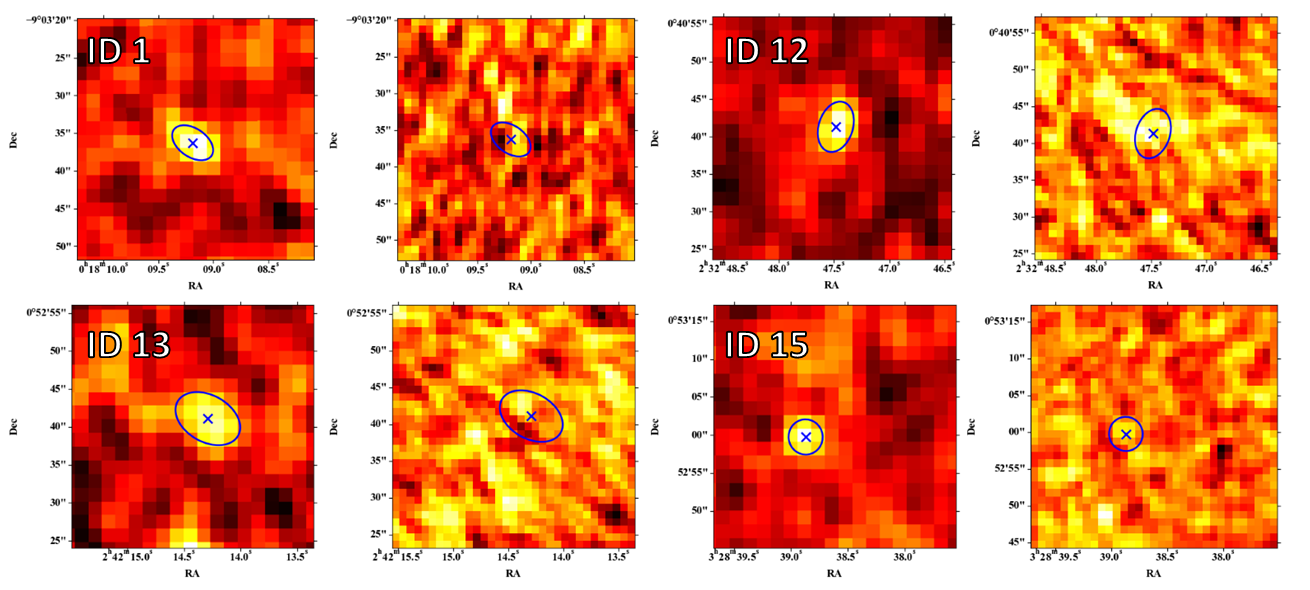}
 \caption{FIRST (left) and VLASS (right) cutout images for four sources from the 9GHz–ND sample that are detected in FIRST but not in VLASS. The fit to each source in FIRST is overlaid on both the FIRST and VLASS images to illustrate the expected source position and extent. These examples highlight the significant flux decline between survey epochs for this subset of variable candidates.}
 %\jme{may redo for higher res images -- not look great when zoom in, see variables.pub}
  %If do 3 sigma, it looks acne-esque (see id 43 in next figure) 4 sigma about the dividing line between what was classifed as source by vla9}
\label{fig:vari_first_vlass}
\end{figure*}

\vspace{.25cm}
\noindent
{\it SNe/SNRs:}
%\label{subsec:var_sne}

We evaluate whether the observed radio variability can be explained by emission from luminous SNe or SNRs, following the approach of \citetalias{reines2020}. We compare the observed 1.4 GHz radio luminosities of the variable sources to the expected maximum luminosities of individual SNe/SNRs based on the empirical scaling between host galaxy SFR and peak SN/SNR radio luminosity derived by \citet{chomiuk2009}. As shown in the top panel of Figure~\ref{fig:bright_sn_var}, all variable sources exceed the predicted luminosities of even the brightest expected SNe/SNRs by more than an order of magnitude, indicating that their emission is unlikely to originate from a single SN event.

Given the modest angular resolution of FIRST, it is possible that the observed radio emission instead arises from the combined contribution of multiple SNRs within the survey beam. To test this scenario, we compare the observed 1.4 GHz luminosities to the expected integrated radio output from a galaxy-wide population of SNRs, again using the relations from \citet{chomiuk2009}. As shown in the bottom panel of Figure~\ref{fig:bright_sn_var}, the variable sources exceed the predicted galaxy-integrated SNR luminosities by factors of $\gtrsim $3. We therefore conclude that neither individual nor collective SN/SNR emission likely account for the observed radio flux densities.

%\ar{(probably want to clarify that the above relations are for normal SNRs/SNe. what about superluminous SNe? add a bit on extreme objects too.)}

%\jme{add stuff about PTF11qcj... include in plot?}
We note, however, that the \citet{chomiuk2009} relations apply to typical SNe/SNRs, and rare superluminous or radio-loud events may reach substantially higher luminosities. For comparison, PTF11qcj, an extreme Type Ic supernova, reached a peak radio luminosity of $\sim 10^{22}~\mathrm{W~Hz^{-1}}$ at 5 GHz \citep{corsi2014, Palliyaguru2019}, despite occurring in a galaxy with a modest SFR of $0.72~M_\odot~\mathrm{yr^{-1}}$. While this example demonstrates that exceptionally luminous radio supernovae exist, such events are exceedingly rare. Indeed, the majority of broad-lined Type Ic SNe are radio faint ($L_\nu \lesssim 10^{12}~\mathrm{erg~s^{-1}~Hz^{-1}}$; \citealt{berger2003, soderberg2006, corsi2016}). %these citations come directly from Palliyaguru intro with same claim
Consequently, although we cannot fully exclude an origin in extreme SNe, the SN/SNR scenario is still disfavored for the variable sample. 

%From corsi2015:  In the past, hundreds of SNe Ib/c have been targeted with the Karl G. Jansky Very Large Array (VLA; Berger et al. 2003a; Soderberg et al. 2006c; Bietenholz et al. 2014) and the fraction of SNe Ib/c associated with GRBs has been constrained to . 1 − 3%
%While rare radio-loud supernovae (e.g., PTF11qcj; \citealt{corsi2016}; \citealt{Palliyaguru2019}) can exhibit double-peaked radio light curves, both radio-loudness and such temporal behavior are uncommon, rendering this explanation improbable for the SF sample.

\begin{figure}
 \includegraphics[width=\columnwidth]{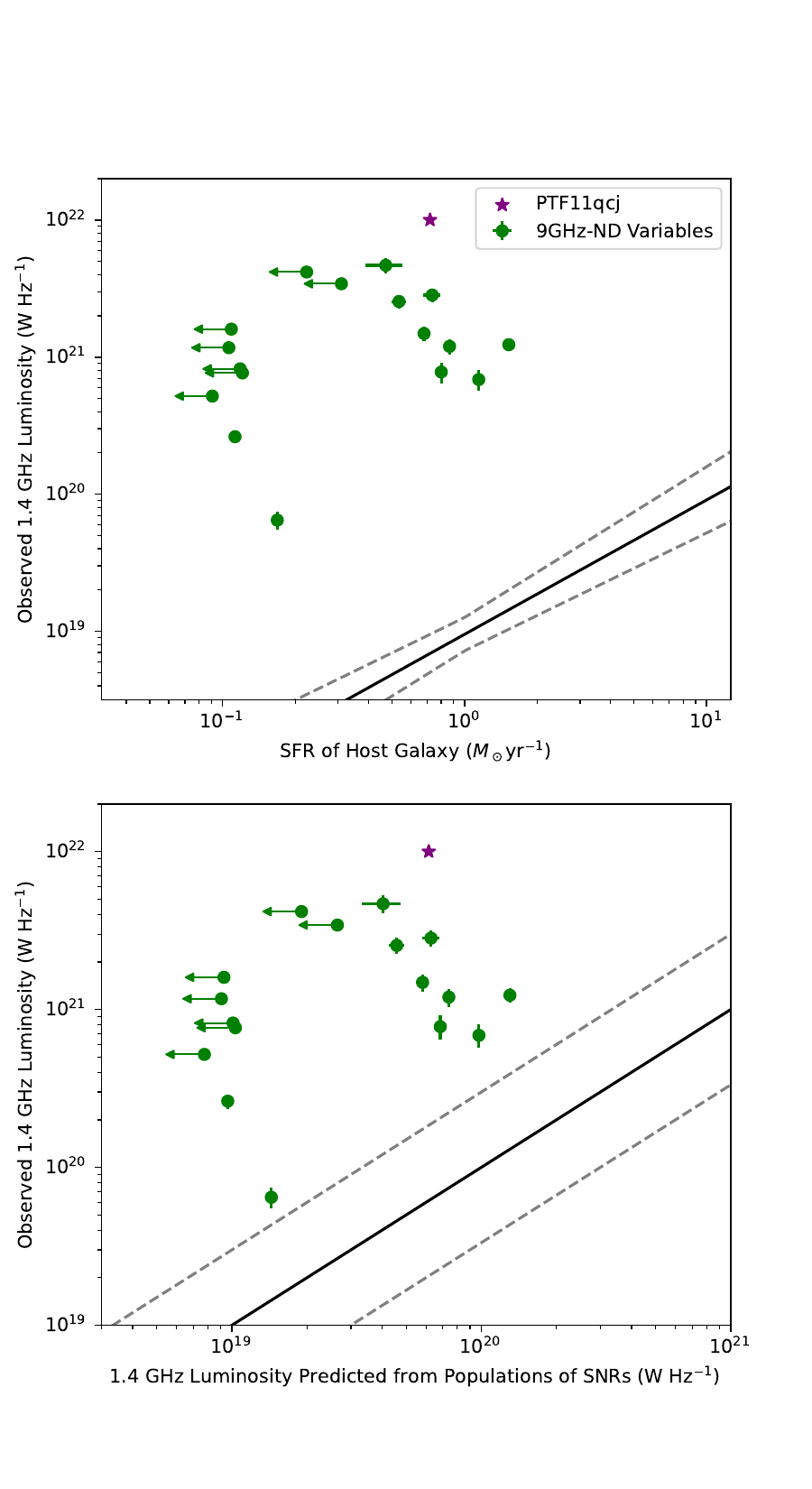}
 \caption{Top: observed 1.4 GHz radio luminosities of variable sources in the 9GHz–ND sample compared to the predicted maximum luminosities of the brightest individual SNe/SNRs in their host galaxies, based on the SFR-dependent relation from \citet{chomiuk2009}. The solid line indicates the expected peak luminosity, and dashed lines represent the scatter due to stochastic sampling. Bottom: observed 1.4 GHz luminosities compared to the expected total radio emission from galaxy-wide SNR populations. The black line denotes the one-to-one relation, with dashed gray lines indicating factors of three above and below this expectation. In both panels, the variable sources significantly exceed the luminosities expected from normal SN/SNR populations. 
 Estimates dependent on upper limits in WISE are indicated with leftward arrows. The luminosities are also compared to PTF11qcj, an extreme Type Ic supernova.}
\label{fig:bright_sn_var}
\end{figure}

\vspace{.25cm}
\noindent
{\it GRBs:}

GRBs represent another potential source of transient radio emission. On-axis GRB afterglows can reach peak radio luminosities of $\sim10^{24}~\mathrm{W~Hz^{-1}}$ \citep{chandra2012}, substantially exceeding those measured for our variable sources. Because GRB radio emission typically evolves on timescales of $\sim$1–2 weeks near peak brightness \citep{pietka2015}, a GRB afterglow could, in principle, produce a detection in FIRST while remaining undetected in VLASS several years later. However, such events are expected to be extremely rare in blind radio surveys. Only $\sim$30\% of GRBs are even detected at radio wavelengths, largely attributed to low sensitivity  \citep{chandra2012}. %and the fact that approximately one in three GRBs are radio-faint \citep{Hancock2013}. 
Given the relatively poor sensitivity of all-sky surveys and the short-lived nature of GRB radio emission,  GRBs are unlikely to contribute significantly to radio-variable source populations in wide-area surveys such as FIRST and VLASS. %We therefore consider GRBs an unlikely explanation for the observed variability.
%,  From laskar 2018: "detailed observations of afterglows in the radio and millimeter have resulted in a low detection rate of about 30%, with sensitivity being the primary challenge (Chandra & Frail 2012; de Ugarte Postigo et al. 2012), %Chandra  Frail 2012 found 31\%  de Ugarte Postigo
%and an even smaller fraction exhibit bright, long-lived radio afterglows comparable to out sample (add references). Moreover, since the radio emission typically evolves on timescales of $\sim$1–2 weeks near peak brightness \citep{pietka2015}, implying a very small duty cycle for detectability. %Given an observed GRB rate of a few hundred per year across the full sky (Wanderman  Piran 2010; Howell  Coward 2013), the expected number of radio-bright GRB afterglows present at any given time is well below unity. 
%Consequently, GRBs are unlikely to contribute significantly to radio-variable source populations in wide-area surveys such as FIRST and VLASS.

\vspace{.25cm}
\noindent
{\it TDEs:}

Tidal disruption events (TDEs; e.g., \citealt{komossa2015}) can also generate transient radio emission as stellar debris accretes onto a central BH. Swift J1644+57, one of the most luminous radio TDEs known, reached a peak radio luminosity of $\sim$$10^{25}~\mathrm{W~Hz^{-1}}$ \citep{eftekhari2018}, again exceeding the luminosities of the variable sources studied here. 
{While TDEs are expected to occur preferentially around the lower-mass BHs typical of dwarf galaxy nuclei} %where stars can be disrupted outside the event horizon rather than swallowed whole by more massive supermassive black holes (Hills 1975).}
\citep{hills1975}, only a small fraction of TDEs produce powerful relativistic jets or detectable radio emission (e.g., \citealt{vanvelzen2018, alexander2020}). Fewer than a few dozen TDEs with confirmed radio emission have been reported to date \citep[e.g.,][]{Goodwin2023}, although recent radio follow-up of 23 TDEs by \citet{cendes2024} indicates that late-time radio emission may be a common, previously under-recognized phase of TDE evolution. Despite this, the overall occurrence rate of radio-bright TDEs remains low, making it unlikely that they contribute a significant fraction of the variable sources in our sample.

{Recent discoveries also highlight a class of unusual, long-lived radio transients that may bridge the gap between classical GRB afterglows and other explosive phenomena. For example, ASKAP J005512.2–255834, a slow, off-nuclear radio transient associated with a star-forming dwarf galaxy, was recently reported by \citet{gulati2026}. This source exhibited a $\sim$20-fold rise in flux density over $\lesssim$250 days and has remained detectable for more than $\sim$1,000 days, with an evolving synchrotron spectrum indicative of an expanding blast wave. Its properties are broadly consistent with either the late-time emission from a long-duration GRB afterglow or a TDE involving an intermediate-mass BH displaced from the galaxy nucleus. The discovery of such a system suggests that rare, long-duration transients may be more diverse than previously appreciated. In this context, some of our sources that are detected in FIRST but not in VLASS could, in principle, represent analogous events observed at later evolutionary stages. However, the apparent lack of current radio detections for our sample, combined with the extreme rarity of such transients, makes this interpretation uncertain without additional multi-epoch or multi-wavelength follow-up.}

%Tidal disruption events (TDEs; e.g., \citealt{komossa2015}) also produce transient radio emission as disrupted stellar debris accretes onto a central BH. The most luminous known radio TDE, Swift J1644+57, reached a radio luminosity of $\sim10^{25}\;\mathrm{W \; Hz^{-1}}$ \citep{eftekhari2018}, exceeding the luminosities of the sources discussed here. While TDEs are more likely to occur around less massive BHs due to their stronger tidal fields \citep{hills1975},  relativistic jet–producing TDEs are rare, constituting only a small fraction of the observed TDE population (e.g., \citealt{vanvelzen2018}; \citealt{alexander2020}). Overall, radoi-emitting TDEs also rare, see abstract of \cite{Goodwin2023}, where les than 20 reported overall, (note the caveat that while at this time there are  fewer than a few dozen TDEs reported with confirmed radio emission to date, Cendes et al. (2024) reports them being more common, fidngin 23 in that paper..., data set indicated 40\% of optical TDEs detected in radio). Nonethelss,  while possible, TDEs are less likely candidates when compared to AGNs.

\vspace{.25cm}
\noindent
{\it AGNs:}

While GRBs and TDEs can, in principle, produce transient radio emission, their rarity makes them statistically unlikely explanations for the majority of the observed variability. By contrast, radio-loud AGNs are a common and persistent population within radio surveys \citep[e.g.,][]{condon1998,  smolcic2017, eberhard2025}. Furthermore, studies have also consistently shown that GHz-frequency radio variability is predominantly associated with AGN activity \citep[e.g.,][]{Hodge2013,Radcliffe2019, Sarbadhicary2021}.

%While GRBs and TDEs are possible explanations for the observed variability, they are relatively rare, compared to AGNs, as previously noted.  %cendes2024: Ubiquitous Late Radio Emission from Tidal Disruption Events
%By contrast, radio AGNs dominate wide-area radio sky catalogs, numbering in the millions in surveys such as FIRST, NVSS, LOFAR, and VLASS. See Condon1998, padovani2017, smolcic2017. AGNs provide the most plausible explanation for our variable sources. Previous studies have demonstrated that radio sources exhibiting variability at GHz frequencies are predominantly associated with AGN activity \citep{carilli2003,bannister2011,thyagarajan2011,frail2012,bell2015,mooley2016}.

Among the seventeen candidate variable sources in the 9GHz–ND sample, five (IDs 13, 37, 45, 63, and 94) are independently identified as radio-excess AGN candidates based on their radio-to-IR emission ratios (see Section \ref{subsec:irrc}). This provides additional evidence supporting AGN activity in these objects. The remaining variable sources, although not separately classified as AGN candidates, are consistent with expectations for low-luminosity or moderately variable AGNs.

%Of the seventeen variable sources, %threw out 104  five (IDs 13, 37, 45, 63, and 94) are independently identified as radio-excess AGN candidates based on their radio-to-IR emission ratios (\S \ref{subsec:irrc}). These are strongly indicating AGN activity. %Among these, IDs 13, 45, and 63 were identified as AGN candidates by both the IRRC and SFR comparison method, ID 37 was only identified as a candidate by the IRRC method, and ID 15 was only identified as a candidate using the SFR comparison method. 
%, with IDs 13, 45, and 63 representing the most robust candidates.

%\ar{(Stick to only talking about variable sources in this section. I'm getting confused. Elaborate more at end of 1st paragraph?)}

{We further assess variability in our sample using data from the Rapid ASKAP Continuum Survey (RACS), which provides observations at $\sim$0.9 GHz, $\sim$1.4 GHz, and $\sim$1.65 GHz (the low-, mid-, and high-frequency catalogs, respectively), obtained over multiple epochs between 2019 and 2022. We find that IDs 29, 75, and 78 have detections in the low- and mid-frequency RACS catalogs, and ID 101 is detected at all three frequencies. The angular resolution of the RACS data is substantially coarser than that of FIRST and VLASS \citep[$\sim25\arcsec$ at low frequency and $\sim10\arcsec$ at mid/high frequencies; see][]{hale2021,duchesne2024,duchesne2025}, and the measured flux densities may therefore include extended or diffuse emission that is resolved out in higher-resolution observations. Nevertheless, the presence of RACS detections in IDs 29, 75, and 78, and 101 indicates that radio emission persists over multi-year timescales. This persistence disfavors an origin in short-lived transients such as GRBs or TDEs, which would likely have faded since the FIRST epoch, making AGN activity the more probable explanation. ID 12, while not detected in the low-frequency catalog, is detected in the mid- and high-frequency catalogs, similarly disfavoring a short-lived transient origin. Although these objects have RACS detections, they still lack compact detections at the higher resolution of VLASS ($\sim2\farcs5$), so we retain their classification as variable sources.}

In sum, while extreme transient events such as SNe, GRBs, or TDEs could plausibly explain a small fraction of strong radio flares, AGNs provide the most likely explanation for the majority of the observed variability. %If any of the variable sources are linked to GRBs or TDEs, they would originate from accreting stellar-mass BHs. 
Multi-frequency and multi-epoch follow-up observations will be essential to distinguish between these scenarios and to further constrain the physical mechanisms driving the variability \citep{nyland2020}.

\subsection{IR-Radio Correlation (IRRC) Parameters}
\label{subsec:irrc}
We next utilize the IR-radio correlation (IRRC) parameter $q$ to help distinguish between radio emission dominated by SF or an AGN.
%To evaluate if the galaxies in our samples are consistent with radio-excess AGN hosts, we calculate the IR-radio correlation (IRRC) parameter $q$ for each the host galaxies. 
The IRRC is a widely used diagnostic for identifying radio-excess AGNs, since star-forming galaxies (SFGs) exhibit relatively tight distributions of $q$ values, with $q$ defined as
\begin{equation}
    q = \mathrm{log}\left[ \frac{\mathrm{L_{IR}}}{3.75\times10^{12} \; \mathrm{L_{1.4\;GHz}}}\right]
    \label{eq:IRRC}
\end{equation}
Here, L$_{\mathrm{IR}}$ is the IR luminosity in W and L$_{\mathrm{1.4\,GHz}}$ is the monochromatic radio luminosity at 1.4 GHz in W Hz$^{-1}$ \citep[e.g.,][]{dejong1985, helou1985}. IR luminosities are typically reported either over the far-IR (FIR; 40-120 $\mu$m) or the total IR (TIR; 8--1000 $\mu$m) wavelength ranges.

%Note that 107 has offset larger than r50, but still well contained within the galaxy
% We note that, owing to the W4 angular resolution of $12\arcsec$, the IR emission is spatially extended on scales larger than the positional offsets between the radio and optical components. Consequently, even for sources in which the radio source is offset from the optical nucleus, the IR and radio measurements probe overlapping regions.}

Radio-excess AGNs are identified as systems with significantly depressed $q$ values, reflecting radio luminosities in excess of those expected from SF. Thresholds of $2$–$3\sigma$ below the mean IRRC are commonly adopted (e.g., \citealt{condon2002,bonzini2013,delmoro2013,delhaize2017}), and \citet{delvecchio2021} demonstrated that a $2\sigma$ criterion optimizes completeness while limiting contamination. 

Using this approach, \citet{eberhard2025} derived a $2\sigma$ threshold of $q_{\mathrm{TIR}} < 1.94$ from a sample of $\sim$7,000 galaxies across the full mass range in the NSA that have radio and IR detections from VLASS (via the CIRADA catalog)} and WISE, respectively. Below this $q_{\mathrm{TIR}}$ value, the radio emission is inconsistent with being SF-dominated. In their analysis, 3 GHz flux densities were converted to 1.4 GHz assuming a spectral index of $\alpha=-0.7$, except in cases where FIRST measurements were available, in which the observed 1.4 GHz fluxes were adopted directly. 

For our 9 GHz ND sample, we compute $q_{\mathrm{TIR}}$
following the same procedure. Since FIRST 1.4 GHz measurements exist for all sources considered here, no spectral-index extrapolation from 3 GHz is required. We estimate $\mathrm{L_{TIR}}$ using the relation from \citet{hao2011}, who reported that $\mathrm{L_{TIR}} \approx  8.33 \mathrm{L_{25\mu m}}$. As in Section \ref{subsec:sf_host_gals}, we estimate $\mathrm{L_{25\mu m}}$ using the W4 band from WISE. {We note that for five sources in the 9GHz-ND sample (IDs 13, 27, 37, 52, and 71), the W4 measurements are reported as upper limits in the WISE catalog. Additionally, six sources (IDs 4, 15, 44, 45, 63, and 98) have radio sources in FIRST that do not align well with their crossmatched dwarf galaxies. In all such cases, the positional offsets between the FIRST radio detections and the nuclei of the dwarf galaxy counterparts exceed the half-light radii of the galaxies. Visual inspection also indicates that the radio emission for these sources is located near the edge of, or beyond, the optical extent of the galaxies. We therefore conclude that these radio sources are unlikely to be physically associated with the corresponding dwarf galaxies. Inspection of the WISE W4 images shows that the IR emission is well-centered on the optical galaxies for these systems, suggesting that the mid-IR flux traces the emission from the host galaxies rather than the offset radio sources. Under these circumstances, the IRRC parameter cannot be meaningfully defined using the measured IR flux densities. Instead, for these objects we adopt the W4 sensitivity limit to compute upper limits on L$_{\mathrm{TIR}}$, which in turn yield upper limits on $q_{\mathrm{TIR}}$.} 

%calculate $q_{\mathrm{TIR}}$ values for the 9GHz-ND sample and 
We compare the $q_{\mathrm{TIR}}$ values to the $2\sigma$ AGN threshold from \citet{eberhard2025}.
We find that eight 9GHz-ND galaxies 
(IDs 4, 13, 27, 37, 45, 52, 63, and 94) have $q_{\mathrm{TIR}}$ values below the AGN threshold (i.e., $q_{\mathrm{TIR}}<1.94$), making them inconsistent with SF.  As such, these galaxies are radio-excess AGN candidates, albeit AGNs without compact radio emission on $\sim 0\farcs25$ scales. {Six of these sources (IDs 13, 27, 37, 45, 52, and 63), have $q_{\mathrm{TIR}}$ values derived using WISE upper limits, and thus represent upper limits themselves. However, even if the true values of $q_{\mathrm{TIR}}$ are lower than these estimates, the objects continue to have  $q_{\mathrm{TIR}}$ values consistent with AGN candidates.} {We also note that none of these sources appear in the \citet{eberhard2025} radio-excess sample because their analysis used only VLASS detections listed in the CIRADA catalog. In our sample, five galaxies (IDs 13, 37, 45, 63, and 94) show no VLASS detections at all (see Section \ref{subsec:variables}), and the remaining sources lack CIRADA catalog entries, but instead have flux densities measured through Gaussian fitting with CASA \texttt{imfit}.}

%\ar{(Are these the same as the ones from your earlier paper? I'm still confused. You are using FIRST values for q rather than VLASS?)}
%IDs 4,13,27,37,45,52,63,
%\jme{No, none of these are the same ones as from the VLASS paper, since none of these were in the CIRADA catalog. Yes, I am using the FIRST values for q, since even in the VLASS paper I converted to 1.4 GHz luminosities to calculate q values.}

\subsection{AGN Candidates}
\label{subsec:agn_cand}
Table~\ref{tab:nd_agn_cand} summarizes all 20 of the 9GHz–ND galaxies exhibiting AGN-like properties, including the variable sources discussed in Section~\ref{subsec:variables} and the radio-excess AGN candidates identified in Section~\ref{subsec:irrc}. Of these, only ID 29 has been previously classified as an AGN in the literature \citep{sacchi2024,eberhard2025b}. In our analysis, however, the radio emission associated with ID 29 is significantly offset from the X-ray source, suggesting that these detections may trace physically distinct systems rather than a single AGN. 

%Table~\ref{tab:nd_agn_cand} summarizes all 9GHz-ND galaxies with AGN-like properties, including the radio-excess candidates from Section \ref{subsec:irrc} and variable sources from Section \ref{sec:variables}. Among these 20 galaxies, only ID 29 has been previously classified as an AGN in the literature \citep{sacchi2024,eberhard2025b}. However, the radio emission in our study is significantly offset from the X-ray source, suggesting the detections may trace distinct physical systems. %We also note that ID 101, while not formally classified as an AGN by the SFR-based diagnostic, lies near the boundary of the AGN plume in the SFR comparison diagram (Figure \ref{fig:first_sfr}) and exhibits significant variability, suggesting a possible AGN origin.  

\begin{figure*}
\includegraphics[width=\textwidth]{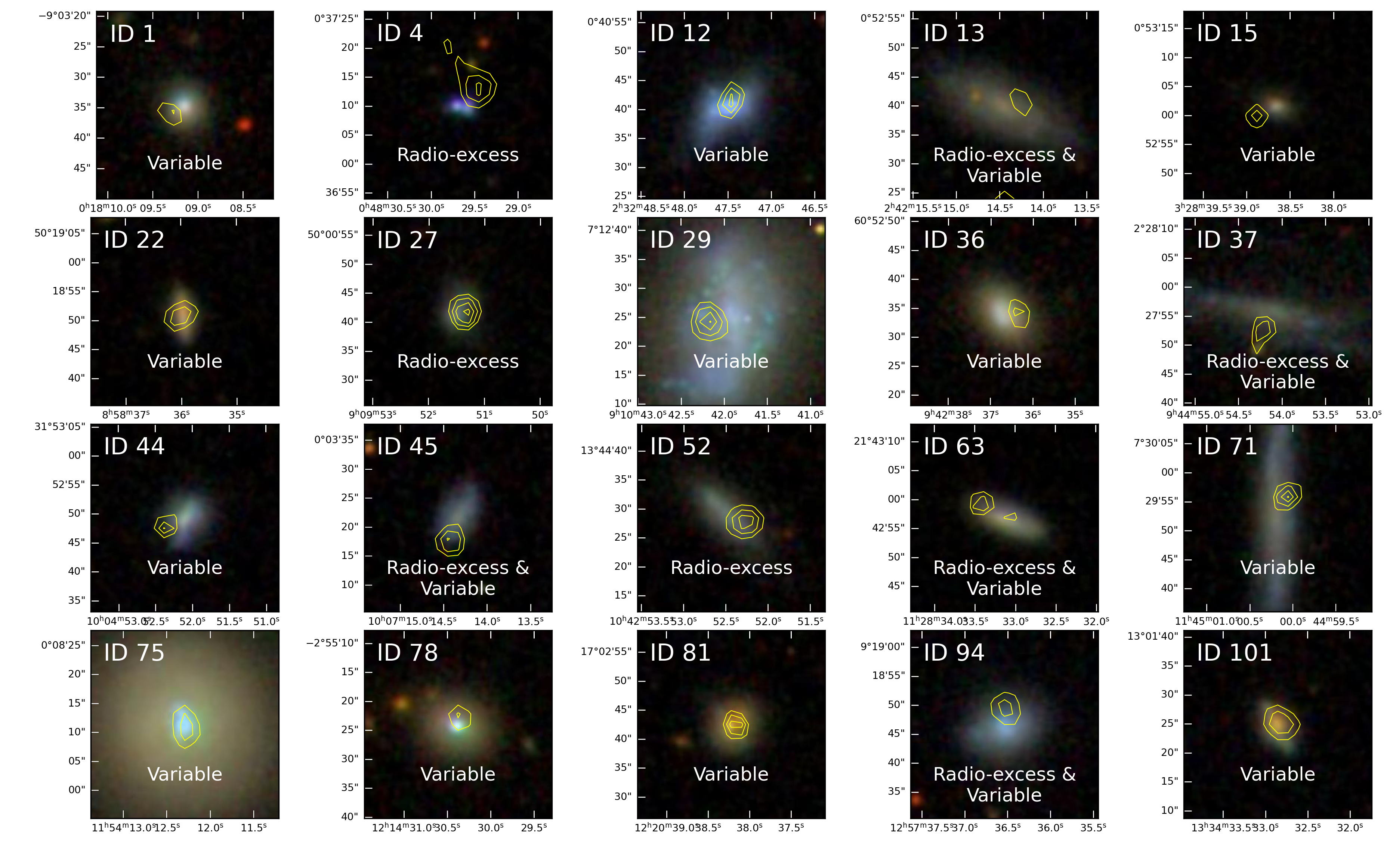}
\caption{FIRST contours (yellow) overlaid on SDSS images for the radio sources with AGN-like properties (i.e., radio-excess or variability). Contours correspond to 3$\sigma$, 4$\sigma$, and 5$\sigma$ above the image mean, where $\sigma$ is the standard deviation of the FIRST background flux. IDs 4, 15, 44, 45, and 63 have FIRST positions that are offset by more than $r_{50}$ from the NSA coordinates for the host dwarf galaxy. These sources are candidates for wandering BHs or offset transient events.}
\label{fig:contours_offset}
\end{figure*}

For the 20 AGN candidates, we examine the relative positions of the FIRST detections and their host galaxies in Figure~\ref{fig:contours_offset}, and list the projected separations between the FIRST radio centroids and the galaxy coordinates from the NSA in Table~\ref{tab:nd_agn_cand}. The sources within IDs 15, 44, 45, and 63 exhibit radio–optical offsets greater than the half light radii of their host galaxies and ID 4 lies at a projected distance exceeding twice the half light radii of its host galaxy. %The radio emission remains spatially coincident with the galaxy light for IDs 15, 44, 45, and 63  (Figure~\ref{fig:contours_offset}), making them plausible candidates for wandering massive black holes or, possibly spatially offset TDEs or GRB afterglows. In contrast, 
The radio source for ID 4 also coincides with an optical source unrelated to the dwarf galaxy, making a background AGN interpretation particularly plausible for this object. %\ar{(what's up with 98 and 107? they are mentioned below but not here? also, it's 94 and 101 in fig 8. I'm confused.)} \jme{IDs 98 and 107 are not AGN candidates; they are neither variable or radio-excess. The only reason I mention them in the paragraph below is because they have offset sources.}

%\jme{Should I even include the following analysis of background sources? I'm not sure it adds much, other than noting that approx. 10\% of our sources overall are expected to be background sources.}
To assess the likelihood of chance superpositions, \citetalias{reines2020} estimated that, among the 186 initial NSA–FIRST matches in their parent sample, $19\pm4$ were expected to be random alignments (e.g., their Figure~1), corresponding to a contamination fraction of $\sim$10\%. Scaling this expectation to the 59 galaxies in our 9GHz-ND subsample implies that $\sim$6 of the FIRST detections may be unrelated background sources. Within the 9GHz–ND sample, there are six sources whose radio positions lie beyond the host galaxy $r_{50}$ (IDs 4, 15, 44, 45, 63, and 98), and are therefore the most likely to be background contaminants. Five of these (IDs 4, 15, 44, 45, and 63) are AGN candidates (as mentioned above), but their classification as AGN candidates in dwarf galaxies should be treated with caution, as they could represent background quasars. We note, however, that crossmatching our sources with the SDSS DR16 list of quasars \citep{lyke2020} %https://www.sdss4.org/dr17/algorithms/qso_catalog/
and the MILLIQUAS catalog \citep{flesch2023}, which contains over one million quasars, yields no counterparts for any of these sources.

%A comparison to their host galaxy images in Figure \ref{fig:vari_sdss_first} reveals that only IDs 27, 75, 87, and 101 are coincident with the galactic nuclei. The other sources, if not GRBs or TDEs, are likely to be either background AGNs or “wandering" BHs. 

%\begin{figure*}
%\includegraphics[width=\textwidth]{contours_ND_AGN.png}
% \caption{SDSS images of the galaxies in the 9GHz-ND sample that have AGN-like properties (i.e., radio-excess and/or variability) The contours levels from the FIRST images are shown in yellow at 3$\sigma$, 4$\sigma$, etc.} %If do 3 sigma, it looks acne-esque (see id 43 in next figure) 4 sigma about the dividing line between what was classifed as source by vla9}
%\label{fig:vari_sdss_first}
%\end{figure*}

%\begin{landscape}
\begin{deluxetable*}
{ccccccc|cc}
%\rotate
\tablecaption{9GHz-ND Galaxies with AGN-like Properties}
\tablewidth{0pt}
\tablehead{
\colhead{ID} &
\colhead{Name} &
\colhead{NSA ID} & 
\colhead{R.A.} &
\colhead{Dec.} & 
\colhead{$r_{50}$} &
\colhead{FIRST Offset} &
%\colhead{SFR Comparison?} &
\colhead{Variable?} &
\colhead{Radio-excess?}
%\colhead{Background/offset} 
}
%see reinesA_MASTER.csv
%see vlass_cirada_astropy_casa.csv for vlass stuff
\colnumbers
\startdata
1 & J0018-0903&21913&4.53817&-9.05993&1.7&0.7&Y&N \\
4 & J0048+0037 & 62105 & 12.12363 & 0.61966 & 1.2 & 3.9 & N & Y\\
12 & J0232+0040&7299&38.19762&0.6781&2.0&0.8&Y&N \\
13 &J0242+0052 &  7427& 40.56008 & 0.87781 & 5.6 & 0.8&Y & Y \\
15 & J0328+0053 &  28952 & 52.16094 & 0.88366 & 1.9&3.9& Y & N \\
22& J0858+5018&16039&134.64973&50.31409&1.9&0.7&Y&N\\
27 & J0909+5000 &  26702 & 137.46454& 50.01164 & 3.0&1.2&N & Y \\
29 &J0910+0712 &  157080 & 137.67445& 7.20699& 8.0&4.3&Y & N \\
36&J0942+6052&12548&145.65249&60.87639&2.2&2.0&Y&N \\
37 & J0944+0227 &  12357& 146.22529 & 2.46544 & 12.5&3.7&Y & Y \\
44 &J1004+3152 & 125334& 151.2173 & 31.88054 & 2.1&3.4&Y& N\\
45 &J1007+0003 &  233& 151.80957 & 0.05577 & 2.6&3.6&Y & Y\\
52 &J1042+1344 &  74657 & 160.71853 & 13.74106&3.8&2.0&N & Y \\
63 &J1128+2142 & 112490 & 172.13786 & 21.71562&2.7&3.8& Y & Y\\
71 &J1145+0729 &  67035& 176.25086 & 7.49778 &9.3&4.6& Y & N \\
75 &J1154+0008 &  1286 & 178.55115 & 0.13663 &6.8&0.6& Y & N\\
78 & J1214-0255 & 3381& 183.62641 & -2.92317 &2.0&0.4& Y & N\\
81&J1220+1702&117748&185.15907&17.04508&2.6&0.1&Y&N\\
94 &J1257+0918 &  77063 & 194.4022 & 9.31286 &3.0&2.7& Y& Y\\
101 &J1334+1301&  70727& 203.6369 & 13.02359 &2.0&0.8& Y & N \\    
\enddata
\tablecomments{Column 1: identification number from \citetalias{reines2020}. Column 2: galaxy name. Column 3: NSA identification number. Column 4: right ascension, in degrees. Column 5: declination, in degrees. Column 6: halflight radius of galaxy, in arcsec. Column 7: offset between FIRST source and galactic coordinates in the NSA, in arcsec. Column 8: is the radio source a variable source, based on a lack of VLASS detections? See Section \ref{subsec:variables}. Column 9: is the radio source a radio-excess AGN candidate, based on the IRRC analysis? See Section \ref{subsec:irrc}.}
\end{deluxetable*}
%\end{landscape}
\vspace{-\baselineskip}\vspace{-\baselineskip}
\label{tab:nd_agn_cand}

%\subsection{ID 104?}
%What the heck is this bright radio source in both FIRST and VLASS that has no clear opticla or IR counterpart? Serendipitous find...

%\begin{figure}
%\includegraphics[width=\columnwidth]{id104.png}
 %\caption{FIRST (top) and VLASS (bottom) images of ID 104, with the elloipse showing the FIRST source is shown. While the radio source wihtin the galaxy is faint  there is a much brighter radio component located outisde the galaxy, that is also prevalent in the VLASS image. \jme{I'm suprised that FIRST counted this as a source, honestly... its super faint... acutally seems brighter in vlass than in first}}
%\label{fig:gr_hist}
%\end{figure}

%\section{Discussion}

\section{Comparisons Across Samples}
\label{subsec:nd_host_gals}
In this section, we examine the host galaxy properties of the SF and 9GHz–ND samples in comparison to the AGN sample from \citetalias{reines2020}, where ID 92 has been moved from the AGN sample to the SF sample. We do so to assess whether the 9GHz–ND sources more closely resemble the SF or AGN populations. We caution that the sample sizes %particularly the AGN subset with only 12 galaxies, 
are modest, and therefore any inferences about broader galaxy populations should be treated with caution.
%In this section, we compare the properties of the host galaxies for the SF, AGN and 9GHz-ND samples to see is there are any clear properties associated with the AGNs or the SF sample and see how the 9GHz-ND sources compares to them (i.e., are the 9GHz-ND sources more AGN- or SF-like?) %\ar{(to learn what?)}. 
%We note that our sample sizes are small, %(particualy the AGN sample which only consists of 12 systems), meaning that applying the results to a larger population should be done with extreme caution.

\begin{figure}
\includegraphics[width=\columnwidth]{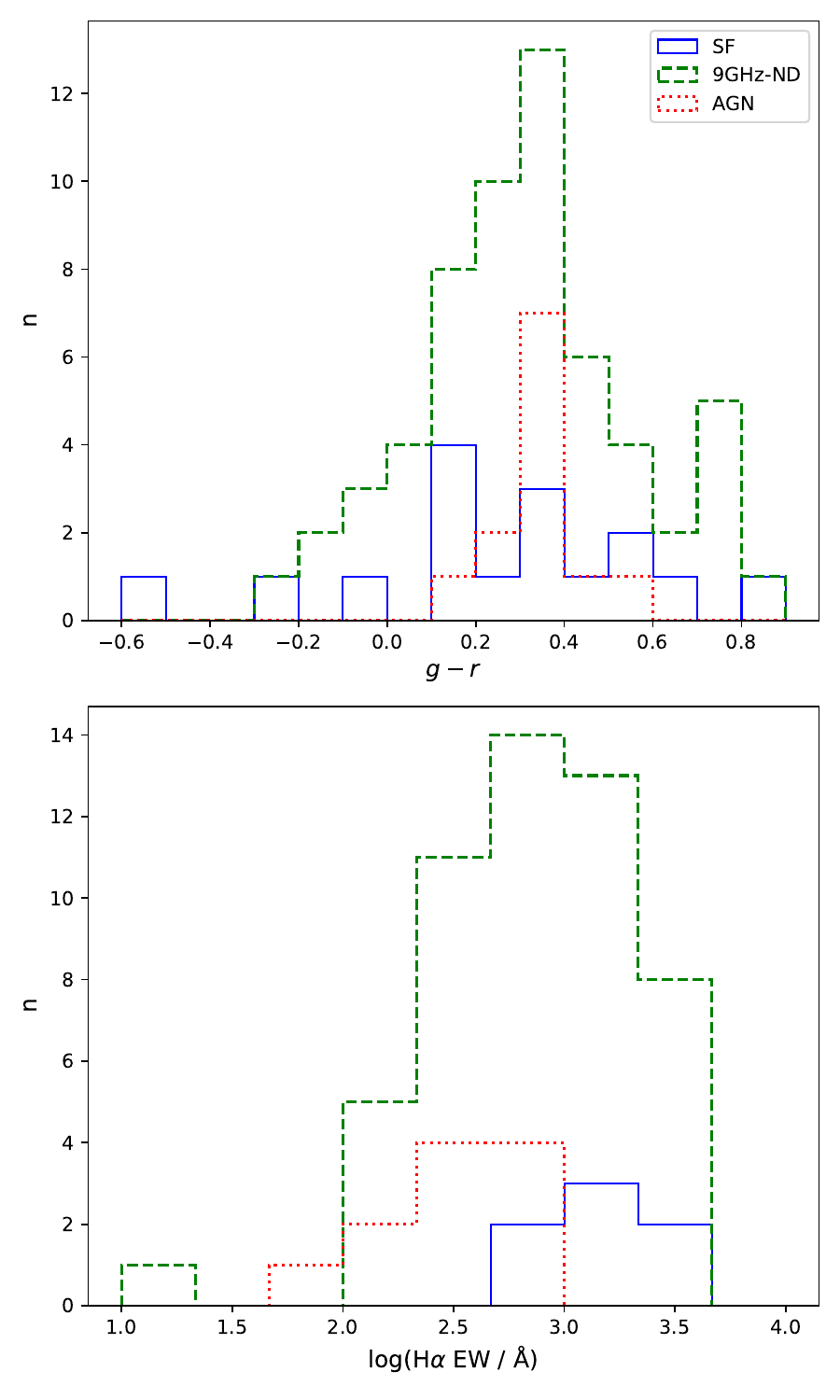}
    \caption{Comparison of galaxy-wide properties between the SF, 9GHz-ND, and AGN samples. Top: $g-r$ color. {Bottom}: log of the H$\alpha$ EWs for those with reliable H$\alpha$ EWs from SDSS spectra. %The 9GHz-ND sample includes eight galaxies with $q_{\mathrm{TIR}}$ values indicative of radio-excess AGNs. 
    }
    \label{fig:sample3}
\end{figure}

The $g-r$ color distributions of the three samples are broadly similar (see the top panel of Figure~\ref{fig:sample3}), with no clear systematic differences. Optical color alone is therefore a poor discriminator between radio-selected SF and AGN activity for the dwarf galaxies in our sample.

%\ar{(Can't get ages from Ha EWs for non-SF things, unless you are assuming Ha only from SF, or argue from BPT. Could just look at Ha EW without converting to ages. Then it's just an observed quantity.)}
A slightly clearer separation emerges in the H$\alpha$ EWs for galaxies with systems with reliable SDSS spectra (see  the middle panel of Figure~\ref{fig:sample3}). AGN hosts all have EWs at or below 636~\AA, and the SF sample has EWs equal to or larger than 589~\AA. The 9GHz-ND sample spans a wide range of EWs, overlapping both the SF and AGN hosts samples, consistent with a heterogeneous population.

\begin{comment}
\begin{figure}
\includegraphics[width=\columnwidth]{gr_hist92.png}
 \caption{$g-r$ color of the 3 samples of galaxies observed with VLA-9.}
\label{fig:gr_hist}
\end{figure}

\begin{figure}
 \includegraphics[width=\columnwidth]{ew_hist92.png}
 \caption{log of the $H\alpha$ EWs of the samples observed with VLA-9, for those with reliable H$\alpha$ EWs from SDSS spectra.}
\label{fig:ew_hist}
\end{figure}

\begin{figure}
\includegraphics[width=\columnwidth]{q_sf_agn_nd.png}
    \caption{$q_{\mathrm{TIR}}$ distribution for the SF, 9GHz-ND and AGN samples. The vertical line marks the threshold at which the galaxies with $q$ values lower than the threshold are strong radio-excess AGN candidates. The 9GHz-ND sample includes eight galaxies with $q_{\mathrm{TIR}}$ values indicative of radio-excess AGNs. }
    \label{fig:q_df}
\end{figure}
\end{comment}

\begin{figure}
\includegraphics[width=\columnwidth]{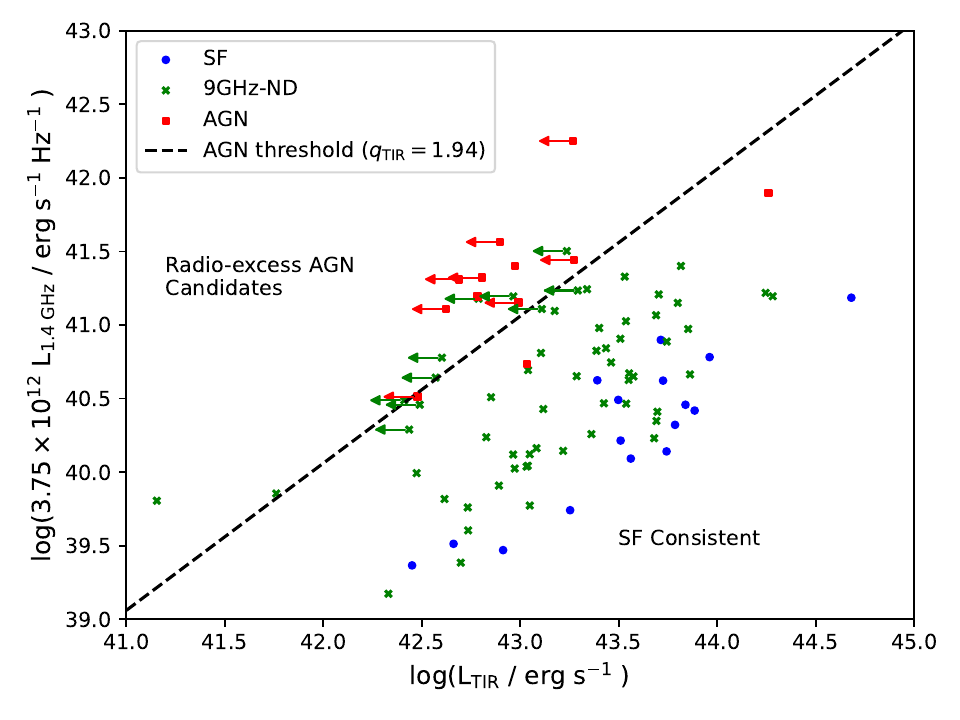}
    \caption{{Comparison of L$_{\mathrm{TIR}}$ and L$_{\mathrm{1.4\;GHz}}$ for galaxies in the SF, 9GHz-ND, and AGN samples. The plotted values are used to compute $q_{\mathrm{TIR}}$ (Equation~\ref{eq:IRRC}) and are shown relative to the radio-excess AGN threshold from \citet{eberhard2025}. Sources with L$_{\mathrm{TIR}}$ derived from WISE upper limits are indicated with arrows.} 
    }
    \label{fig:q_sample3}
\end{figure}

We calculate $q_{\mathrm{TIR}}$ for the AGN, 9GHz-ND, and SF samples and compare them to the 2$\sigma$ AGN threshold from \citet{eberhard2025}. 
{ Figure~\ref{fig:q_sample3} presents the corresponding L$_{\mathrm{TIR}}$ and L$_{\mathrm{1.4\;GHz}}$ values used to derive $q_{\mathrm{TIR}}$ (Equation~\ref{eq:IRRC}), compared to the adopted AGN threshold.} As expected, all of the SF galaxies exhibit $q_{\mathrm{TIR}}$ values consistent with SF. Of the twelve galaxies in the AGN sample (where the AGN sample no longer includes ID 92), nine are inconsistent with SF. The remaining three AGNs (IDs 6, 26 and 82) show $q_{\mathrm{TIR}}$ values consistent with SF, although we note that IDs 26 and 82 are independently identified as AGNs via X-ray and optical diagnostics (see \citealt{reines2013}; \citealt{molina2021}). %This illustrates the limitations of using low-resolution radio data alone, since the IRRC method may exclude genuine AGNs.
As previously noted, eight 9GHz-ND galaxies have values below the threshold. The remaining 9GHz-ND galaxies have a skew towards $q_{\mathrm{TIR}}$ values more consistent with the SF sample.

\begin{figure}
 \includegraphics[width=\columnwidth]{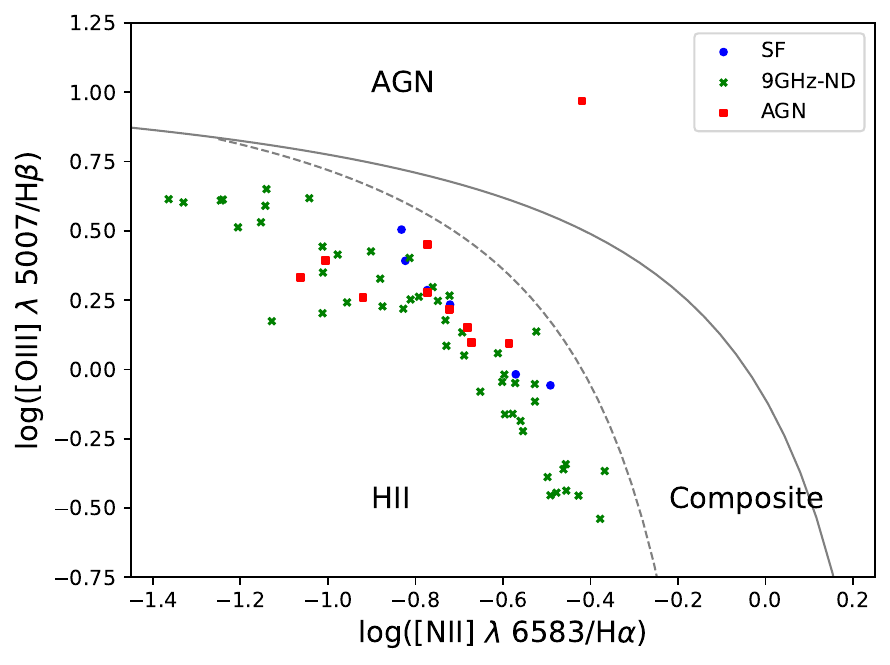}
 \caption{BPT diagram of the galaxies observed by VLA-9 that had SDSS spectral line fits. Divisions between the regions of the diagram come from the classification scheme in \citet{kewley2006}. Only 1 galaxy (ID 26, in the AGN sample) has optical emission lines consistent with an AGN. }% -- rest are HII. This is from idl files }
\label{fig:bpt}
\end{figure}

%\subsection{Infrared and Optical Analysis}
%\label{subsec:nd_multi_wave}
We further compare the 9GHz-ND sample to the AGN and SF sample using multi-wavelength AGN diagnostics. Comparing line strengths from the SDSS spectra in a Baldwin, Phillips \& Terlevich (BPT) diagram \citep{baldwin1981}, we identify only a single object (ID 26) as an AGN in the optical regime , with all other galaxies (including most radio-selected AGN candidates) falling in the HII region (Figure~\ref{fig:bpt}). {ID 26 was previously reported as an optical AGN by \citet{reines2013} (their ID 9), as noted in \citetalias{reines2020}}.

{HeII $\lambda$4686 emission lines have also been shown to be effective at detecting the presence of an AGN in SF-dominated galaxies, and diagnostic diagrams such as HeII $\lambda$4686/H$\beta$ vs. [NII] $\lambda$6584/H$\alpha$ have been used to separate SF from AGN-powered systems \citep[e.g.,][]{shirazi2012}. For our sample, robust measurements of the HeII $\lambda$4686 line are available for only a few galaxies. Following \citet{shirazi2012}, we adopt a signal-to-noise ratio threshold of 5.5 for robust He II $\lambda$4686 measurements. In the available SDSS spectra for our sample, however, this criterion is met only for IDs 7 and 92 in the SF sample and ID 6 in the AGN sample. Even for these galaxies with robust HeII measurements, their line ratios place them within the star-forming region of the HeII $\lambda$4686/H$\beta$ vs. [NII]$\lambda$6584/H$\alpha$ diagram. All other galaxies have either undetected or low signal-to-noise HeII $\lambda$4686 emission, and therefore we cannot place meaningful constraints on their HeII luminosities or line ratios. Consequently, we do {not} find any galaxies that exhibit HeII emission higher than that expected from SF alone. We note that detecting HeII lines at the levels required for such an analysis generally requires higher signal-to-noise or deeper spectroscopy than is available in the current SDSS archival data.}

\begin{figure}
 \includegraphics[width=\columnwidth]{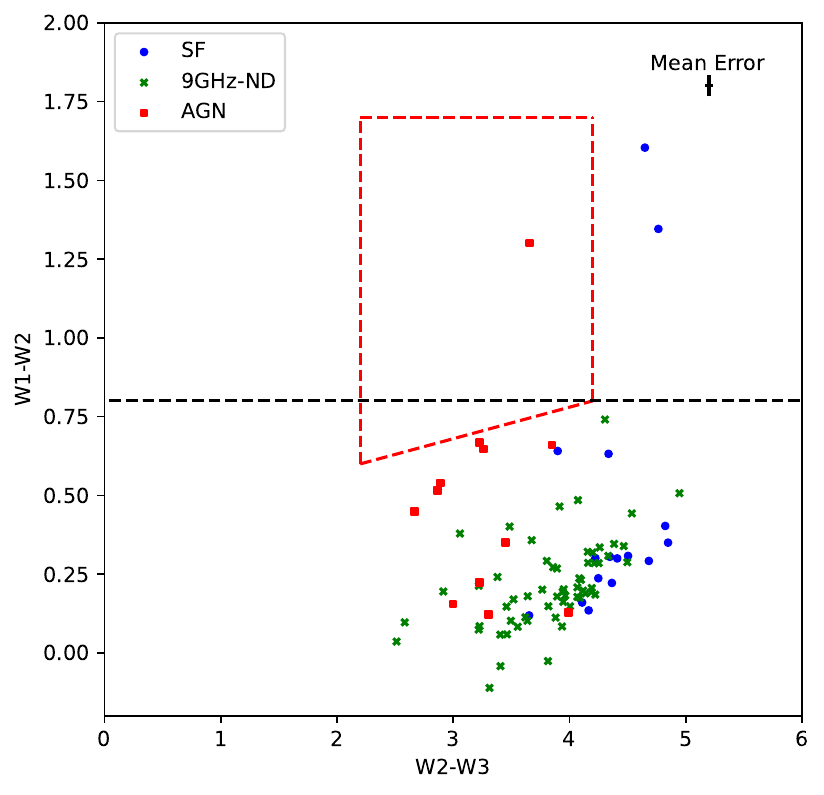}
 \caption{WISE colors of the galaxies studied observed with VLA-9. The \cite{stern2012} cutoff is shown in black and the \cite{jarrett2011}
selection box is shown in red. We find that 2 galaxies in the AGN sample (IDs 26 and 65) and 2 galaxies in the SF sample (IDs 38 and 92) have colors consistent with mid-IR AGNs in at least one of the criteria.}
\label{fig:wise_color_color}
\end{figure}

Mid-IR diagnostics using WISE colors similarly provide limited evidence for AGN activity. Only one source (IDs 26; the optical AGN mentioned above) lies within the AGN selection region defined by \citet{jarrett2011}. {This object was previously classified as an mid-IR AGN in \citet[][their ID 2]{latimer2021}, as noted by \citet{sturm2026}}. Two SF galaxies (IDs 38 and 92) also satisfy the \citet{stern2012} criterion, despite being independently classified as star-forming in the literature (Figure~\ref{fig:wise_color_color}). This is in line with \citet{hainline2016}, who found that dwarf starburst galaxies can mimic the mid-IR colors of AGNs, especially if only using a W1$-$W2 cut. %They also found that the dwarf galaxies with the reddest mid-IR colors have the youngest stellar  highest SFRs, which is consistent with out findings, since the SF sample has red mid-IR colors in Figure~\ref{fig:wise_color_color}. 
In contrast, no 9GHz-ND galaxies meet the mid-IR AGN selection criteria.

%from BPT_idls92.ipynb

%most live in spiral/staburst reiongs, with one weird 9GHz-ND guy (107) off the left, in the eliptical region

%figure from wise_plots.ipynb

{We also consider the expected X-ray emission from our sample to assess whether it is consistent with SF alone. None of the galaxies are detected in available Chandra observations, and only one source (ID 29) has an X-ray counterpart in eROSITA All-Sky Survey. As discussed in Section \ref{subsec:agn_cand}, ID 29 has previously been identified as an AGN candidate; however, the X-ray emission is spatially offset from the radio emission, suggesting that the two may not be physically associated. To contextualize these non-detections, we estimate the X-ray luminosity expected from high-mass X-ray binaries (HMXBs) based on the SFRs of the galaxies. We adopt the scaling relation from \citet{lehmer2019}, $L_{\mathrm{HMXB}} = \beta \; \mathrm{SFR}$, with log($\beta$)$\approx 39.71$ (their Table 4, full sample). For the 9GHz-ND sources, the predicted luminosities span $L_{\mathrm{2–10 keV}} \sim 10^{37}$–$10^{40.5}~\mathrm{erg~s^{-1}}$, corresponding to fluxes of $\sim 3\times10^{-16}$ to $5\times10^{-14}~\mathrm{erg~s^{-1}~cm^{-2}}$. These fluxes lie at or below the sensitivity limits of current X-ray observations. In particular, they are generally below the detection threshold of wide-area surveys such as eROSITA, which has a median flux limit (at 50\% completeness)  of $\sim 5\times10^{-14}~\mathrm{erg~s^{-1}~cm^{-2}}$ \citep{merloni2024}. These fluxes would also only be accessible with relatively deep pointed observations from Chandra, which naturally explains the absence of X-ray detections in the existing data. Targeted X-ray observations will therefore be essential to better distinguish between SF-powered emission and potential AGN activity.}

Overall, the 9GHz-ND sample exhibits a wide range of colors, EWs, and $q_{\mathrm{TIR}}$ values, suggesting it comprises a mixture of SFGs and AGN hosts. Additionally, the limited overlap between radio, optical, mid-IR, {and X-ray} AGN diagnostics underscores the complementarity of radio observations, which can identify accreting BHs in systems where traditional optical and mid-IR indicators are weak or absent.

\section{Conclusion}
\label{sec:conclusion}
%\jme{maybe do bullet points of the following: SF :: high sigma and o-star count, mostly thermal, ND :: IRRc, variables, offset sources}
We analyzed dwarf galaxies with radio detections in the FIRST survey that were subsequently observed with high-resolution 9 GHz VLA imaging in \citetalias{reines2020}. Our study focused on systems not previously classified as AGN candidates, including galaxies exhibiting compact radio emission consistent with SF as well as those lacking compact nuclear sources.
 %A complementary analysis of the AGN candidates identified in this sample is presented in Sturm et al. (2025, in preparation).
 
For galaxies with compact radio emission consistent with SF, we use data from VLASS to confirm that the emission is largely consistent with thermal processes arising from extreme HII regions. Interpreting the emission as thermal implies exceptionally high SFR surface densities up to $\Sigma_{\mathrm{SFR}}$ = $850\,M_\odot\,\mathrm{yr^{-1}\,kpc^{-2}}$, %\ar{(give some numbers)}
with all but one radio source exceeding the galaxy-wide SFR surface density limit reported by \citet{meurer1997}. These galaxies represent systems with ongoing, concentrated SF and exhibit properties consistent with $\sim 400-90,000$ equivalent O-type stars. % \ar{(how many O stars...?)} %broadly consistent with BCDs, making them plausible local analogs to high-redshift starbursts \ar{(no basis presented for BCDs/high-redshift, need to include some literature/REFs)}. 
%However, additional observations, particularly those that further constrain the spectral indices and H$\alpha$ emission, would be useful to robustly determine the thermal fractions and confirm this interpretation. 

Among galaxies lacking compact radio emission in the 9 GHz observations from \citet{reines2020}, we identify eight systems exhibiting diffuse radio luminosities in excess of those expected from their IR luminosities, consistent with the presence of radio-excess AGN activity. Five of these sources, along with an additional 15 galaxies, are detected in FIRST but undetected in VLASS, showing evidence for significant radio variability on multi-year timescales. The observed luminosities of these objects are inconsistent with emission from SNe or SNRs, supporting an interpretation of AGNs or other BH-related phenomena such as TDEs or GRBs. In total, 20 galaxies in our sample exhibit radio properties suggestive of AGN activity: eight are radio-excess and 17 are variable, with five sources falling in both categories. %The five sources that are both radio-excess AGN candidates and strongly variable constitute especially promising targets for high-resolution, multi-wavelength follow-up. %Additionally, 
Five AGN candidates are significantly offset from their host galaxy nuclei, raising the possibility of wandering black holes or background contaminants.

This work highlights the diversity of compact radio phenomena in dwarf galaxies and demonstrates that extreme SF and accretion-driven activity both are prominent in this low-mass regime. %\jme{The galaxies in the SF sample have extreme star-forming environments that may provide a valuable laboratory for investigating conditions analogous to those prevalent during more intense epochs of cosmic SF.} 
{The galaxies in the SF sample host intense, compact star-forming regions that are consistent with the types of dense, clustered environments thought to give rise to SSCs. As such, these systems provide nearby laboratories for investigating the physical conditions that drive an extreme clustered mode of SF that was prevalent in the earlier Universe, as well as the formation of the most massive stars.} The AGN candidates in the 9GHz-ND sample provide insight into the radio AGN population in dwarf galaxies, with potential long-term implications for constraining BH seeding and growth scenarios. Future high-sensitivity, multi-frequency radio campaigns, combined with spatially resolved optical spectroscopy and X-ray observations, will be useful for confirming the nature of these sources and characterizing the interplay between SF and BH accretion in low-mass galaxies.

%From megan: "I think the conclusion is good. Great job on the paper! I think overall the analysis/discussion is solid. I think it just needs to be a bit clearer on the goals and what exactly you are doing that is new versus what has been done in the past, if that makes sense. I will say it might be good to have a general discussion of SF in dwarf galaxies/high z galaxies here in the conclusion as well"

% As next-generation radio surveys expand both sensitivity and time-domain coverage, samples such as this will serve as a critical benchmark for understanding the faint end of the AGN population and the role of BHs in galaxy evolution at low masses.

%\jme{Add in a concluding paragraph explaining why this work is important.}
%\ar{Need more quantitative details in summary/conclusions. Also need a better ending.}

%Our findings contribute to the broader effort to understand the onset and evolution of BH activity and star formation in low-mass galaxies. 
%By studying identifying new radio-excess and variable AGN candidates in the dwarf galaxy regime, this work expands the census of accreting BHs at low masses and highlights the diversity of radio-emitting phenomena in such systems. These results also inform ongoing efforts to understand the coevolution of BHs and SF across cosmic time, helping to connect local dwarf galaxies with their high-redshift counterparts.

\section*{Acknowledgments}
We thank the anonymous referee for their constructive feedback which helped improve the manuscript. A.E.R. gratefully acknowledges support for this work provided by NSF through CAREER award 2235277. We also thank Dr. Megan Sturm for valuable discussions that improved the clarity of the manuscript.

\bibliography{bib}{}
\bibliographystyle{aasjournal}

\end{document}